\documentclass[acmsmall]{acmart}

\AtBeginDocument{%
  }

\setcopyright{acmlicensed}
\copyrightyear{2027}
\acmYear{2027}
\acmDOI{XXXXXXX.XXXXXXX}
\acmJournal{PACMHCI}
\acmNumber{GROUP}

\begin{document}

\title[Decolonial Discourse in Postcolonial Contexts]{Decolonial Discourse in Postcolonial Contexts: How YouTubers Negotiate Audience Tensions, Platform Governance, and State Influence}

\author{Dipto Das}
\authornote{The author was affiliated with the University of Colorado Boulder while working on this study as part of his dissertation.}
\email{dd749@cornell.edu}
\affiliation{%
  \department{College of Computing and Information Science}
  \institution{Cornell University}
  \city{Ithaca}
  \state{New York}
  \country{United States}
}

\author{Bryan Semaan}
\email{bryan.semaan@colorado.edu}
\affiliation{%
  \department{Department of Information Science}
  \institution{University of Colorado Boulder}
  \city{Boulder}
  \state{Colorado}
  \country{United States}
}

\begin{abstract}
  Decolonial discourse on online platforms is often framed in terms of creator motivations and expressive possibilities. In this paper, we examine what it takes to sustain such discourse under layered sociotechnical constraints. Drawing on semi-structured interviews with YouTubers engaging in Bengali decolonial discourse, we analyze how audience publics, platform governance, and state influences shape what becomes sayable, visible, and viable. We show how fragmented postcolonial identities among audiences produce legitimacy policing, harassment, and coordinated backlash, requiring ongoing relational labor from creators. At the platform level, differential monetization, opaque moderation, and copyright regimes reorganize which publics are economically viable and reinforce existing hierarchies. Further, intermediaries such as multi-channel networks mediate regulatory pressure, introducing political risks and constraints on participation. In response, content creators engage in strategies of negotiation, including boundary work, infrastructural improvisation, and multi-platform distribution. Overall, our findings highlight the layered dynamics of decolonial discourse in postcolonial contexts and the continuous work required to sustain it in platformed environments.
\end{abstract}

\ccsdesc[500]{Human-centered computing~Empirical studies in collaborative and social computing}

\keywords{YouTube, Bengali, Decolonial, Postcolonial, Identity}


\maketitle
\section{Introduction}
Digital platforms have become important sites for articulating, contesting, and reimagining identities~\cite{haimson2016digital, morioka2016identity, das2022collaborative}. Video-sharing platforms such as YouTube, in particular, enable creators to challenge dominant historical narratives, foreground local perspectives, and reinterpret cultural meanings through multimodal forms of expression~\cite{chen2023my, lu2019feel}. Such practices can contribute to critical engagements with colonial histories and their continuing effects, seeking to reclaim cultural identity, epistemic authority, and the capacity to define communities on the terms of local, native, and Indigenous peoples--which prior scholarship dubs decolonial discourse~\cite{das2024reimagining, das2022collaborative}. Instead of treating particular video genres as inherently decolonial, decolonial computing scholars emphasized the role of a particular video in revisiting contested histories, challenging exclusionary narratives, foregrounding marginalized cultural perspectives, and fostering understanding among communities separated by postcolonial borders~\cite{das2024reimagining}. However, producing such content does not ensure that it can remain visible, economically viable, politically safe, or socially legitimate. Our analysis shifts attention from the purposes of these formats to the social, infrastructural, and political work required to sustain them over time. 


We examine the sustenance of decolonial discourse in the Bengali\footnote{Bengali is the exonym for both the ethnolinguistic group of people (endonym Bangali) and the language (endonym Bangla).} geocultural context, which spans Bangladesh and parts of India and is shaped by successive colonial and postcolonial transformations. British colonial rule and the 1947 Partition divided Bengal along religious and territorial lines, placing Hindu-majority West Bengal within India and Muslim-majority East Bengal within Pakistan~\cite{marshall2006bengal, sen2018decline, pandey2001remembering}. The political, linguistic, and economic marginalization of East Pakistan subsequently culminated in the 1971 Liberation War and the formation of Bangladesh~\cite{van2020history}. These transformations did not allow a unified postcolonial Bengali public to emerge; instead, they fractured the community in terms of language, religion, nationality, region, and migration~\cite{chatterjee1993nation}. These dynamics persist across the Indian subcontinent, where Bengali Muslims are often labeled ``illegal" migrants and Hindu Bengalis in Bangladesh ``Indian agents," reflecting tensions around language, religion, and national belonging~\cite{maitra2026india, hrw2025india, ahmed2025pluralism}. As a result, creators engaging in decolonial discourse operate within a transnational environment where audiences are not unified and where legitimacy of cultural belonging is composed of multiple, overlapping publics shaped by postcolonial institutions, historical divisions, and geopolitical experiences~\cite{anderson2006imagined, warner2021publics}. 

Existing research in human-computer interaction (HCI) and computer-supported cooperative work (CSCW) has examined how online platforms support identity work and enable marginalized communities to challenge dominant narratives~\cite{dosono2018identity, das2022collaborative}. Complementary work has shown how platform governance structures participation through monetization systems, moderation practices, and algorithmic visibility~\cite{kumar2019algorithmic, jiang2023trade}. Building on these strands of research, we examine how creator-audience interactions, platform governance, and broader sociopolitical conditions become entangled in shaping participation in postcolonial contexts~\cite{irani2009postcolonial}. We argue that decolonial discourse on online platforms is a situated collective practice involving ongoing coordination among creators, heterogeneous audience publics, platforms, intermediaries, collaborators, and political institutions. We ask: \emph{How do Bengali YouTubers negotiate audience tensions, platform governance, and state-linked political pressures to sustain decolonial discourse across transnational postcolonial publics?}


To answer this question, we draw on semi-structured interviews with 15 YouTube creators residing in Bangladesh, India, and Pakistan. We analyze their accounts of interacting with audiences, navigating monetization, moderation, and copyright systems, and responding to political and regulatory uncertainty. Our analysis examines both the constraints creators encounter and the relational, rhetorical, infrastructural, and risk-management practices they employ to remain active. This paper contributes to HCI and CSCW research on collective participation and platformed cultural production in three ways. First, we extend relational labor and online identity scholarship by showing that, in a transnational postcolonial setting, building relationships with audiences becomes historically situated boundary work through which creators establish cultural and political legitimacy. Second, we conceptualize the \emph{selective sustainability} of decolonial discourse to explain how audience boundary policing, unequal platform economies, and state-linked political risk recursively shape which narratives remain socially acceptable, economically viable, and politically survivable. Third, we extend trust and safety scholarship by showing how context-specific anticipatory risk across collaborators, social networks, and production infrastructures shapes what content reaches the platform. Together, these contributions foreground the relational, economic, and political conditions that shape the scope and feasibility of decolonial discourse. Similar to most qualitative research~\cite{leung2015validity}, this paper aims not to produce generalizability but rather to examine how decolonial discourse is negotiated within a specific sociotechnical and postcolonial context.

\section{Literature Review}
In this section, we discuss how colonial histories produce contested publics among Bengali communities, where creators' identity claims are interpreted, authenticated, and policed by heterogeneous audiences. Then, we conceptualize decolonial discourse as a platformed cultural production through which creators revisit histories, represent cultural practices, and negotiate collective identities. Next, we describe how platform governance and broader regulatory environments structure the visibility and viability of this work. Together, these perspectives underscore the importance of our study in understanding how decolonial discourse is sustained on YouTube.

\subsection{Postcolonial Identity Manifested as Contested Publics}
Identity is not fixed or internally possessed but relational, negotiated, and performed through social interaction~\cite{erikson1968identity, goffman1978presentation, gecas1982self}. It is also shaped by collective affiliations, through which people define themselves and are recognized by others based on perceived membership in social groups~\cite{snow2001collective, tajfel1974social}. In postcolonial contexts, these negotiations are structured by colonial systems that privileged simplified categories such as religion, ethnicity, language, and nationality, often overlooking more complex and overlapping forms of belonging~\cite{lugones2016coloniality, chatterjee1993nation, nandy1989intimate}. The Bengali context illustrates how these classifications continue to shape collective identity. Communal riots, subsequent partition, and nation-state formation produced enduring fractures across religious, linguistic, and national lines~\cite{chatterji2002bengal, chatterjee1993nation, pandey2001remembering}. Beyond dividing populations geographically, these processes reshaped how people negotiate belonging across sometimes-competing categories, such as Bengali, Indian, Bangladeshi, Pakistani, Hindu, or Muslim~\cite{das2022collaborative, madan1972two}. Caste, migration histories, agrarian--urban differences, and linguistic practices further complicate these negotiations~\cite{sen2018decline}. In short, native ethnocultural identity that is the objective of decolonial discourse is deeply situational and contested.

These dynamics become especially visible within publics, where identity is not only expressed but also evaluated. Building on imagined communities~\cite{anderson2006imagined}, publics comprise people who orient themselves toward others whom they may never directly encounter but nonetheless imagine as part of a shared audience~\cite{warner2021publics}. In digitally mediated publics, audiences actively interpret, respond to, and regulate discourse, helping determine which expressions are considered legitimate, authentic, or acceptable~\cite{lindtner2011towards, baym2012socially, marwick2011tweet}. Such judgments can involve boundary policing, through which speakers whose identities or perspectives do not align with dominant expectations are questioned, challenged, or excluded~\cite{papacharissi2015affective, marwick2013status}. For content creators, audience engagement therefore involves more than attracting viewers. Creators perform relational labor to build trust and maintain relationships while navigating expectations concerning language, religion, nationality, gender, and cultural legitimacy~\cite{baym2015connect}. These expectations may reproduce hierarchies that privilege particular practices, aesthetics, and forms of expression while marginalizing others~\cite{hall1989cultural}. As colonization categorized and hierarchized Bengali identities~\cite{sinha2017colonial, spivak2003can}, we conceptualize Bengali YouTubers' audiences not as a unified public but as multiple, overlapping publics shaped by different historical narratives and sociopolitical conditions. This framing allows us to examine how creators negotiate competing claims about identity, legitimacy, and representation within transnational decolonial discourse.

\subsection{Decolonial Discourse as Platformed Cultural Production}
Decolonial discourse involves critically engaging with colonial histories and confronting their enduring effects on identity, culture, and systems of knowledge, while actively reclaiming autonomy over self-definition~\cite{fanon2008black, fanon2007wretched, laenui2000processes}. In doing so, people construct, negotiate, and maintain understandings of themselves and their communities as well as ways of knowing beyond colonial frameworks~\cite{laenui2000processes}. Online communities have emerged as important sites for such discourse, where individuals and groups engage in critical conversations about historical and contemporary structures of power, identity, and marginalization~\cite{das2023studying}. In these spaces, people challenge dominant narratives, foreground local perspectives, and articulate alternative understandings of culture, history, and belonging~\cite{das2023decolonization, das2022collaborative}. Such discourse can overlap with identity work~\cite{ibarra2010identity}, through which people construct, negotiate, and maintain understandings of themselves and their communities.

Online platforms can facilitate such work by providing opportunities for self-expression, community formation, and collective sensemaking~\cite{lingel2014city, munoz2022platform, dosono2017exploring}. Marginalized communities, in particular, have used these spaces to negotiate and affirm their identities, build solidarity, and contest exclusionary norms~\cite{dym2019coming, al2010blogging, dosono2018identity, dosono2020decolonizing}. In the context of postcolonial societies, such practices often take the form of collaborative efforts to reinterpret history, challenge hegemonic narratives, and reconstruct collective identities in ways that reflect local experiences and perspectives~\cite{das2022collaborative, das2024reimagining, mukhongodecolonizing}.

Much of the existing work on decolonial discourse and identity negotiation has focused on text-based or discussion-oriented platforms such as Reddit and Quora, where users engage in conversational exchanges and collaborative knowledge production~\cite{das2022collaborative, dosono2018identity}. These studies have highlighted how users collectively construct meaning, negotiate identity boundaries, and develop shared interpretations of sociopolitical issues. However, video-sharing platforms such as YouTube introduce distinct affordances for decolonial discourse and identity work. As a medium, user-generated video enables creators to present embodied and multimodal expressions of identity, through combinations of language, accent, dress, visual aesthetics, and narrative structure~\cite{burgess2018youtube, lange2007publicly, cha2007tube}. Thus, it can support the formation of online publics around cultural practices, sociopolitical issues, and regional identities, enabling users to connect with dispersed audiences and participate in broader public discourse through platforms such as YouTube~\cite{milliken2008user-canada, milliken2008user-sphere, askanius2011online, askanius2014video}, TikTok~\cite{simpson2021you}, and Vine~\cite{mcroberts2016viewers, barta2021constructing}. Recent work demonstrated how marginalized and non-Western communities use video-based platforms like Douyin and Kuaishou to document and promote cultural practices, thereby contributing to their sustainability and visibility~\cite{lu2019feel, chen2023my}.

The closest work to this paper is Das and colleagues' study~\cite{das2024reimagining}, which examines YouTubers' motivations and the types of videos they produce to engage in decolonial discourse by reclaiming cultural narratives and reimagining communities. Their findings show how creators across Bangladesh, India, and Pakistan engage in practices such as travel vlogging, social interviews, reaction videos (videos in which creators record and share their real-time responses to existing media content), political explainers, satire, and YouTube journalism to negotiate fragmented Bengali local and native identities through discussing local culture, history, and politics. Its transnational context highlights how such video-based participation can make identity claims more susceptible to scrutiny, contestation, and policing by diverse audiences. Overall, online communities do not serve as uniformly empowering spaces. Scholars have documented how digital platforms can reproduce and amplify existing social hierarchies, including those based on race~\cite{klassen2021more, klassen2022black}, gender~\cite{ghoshal2020toward, sharma2017analyzing}, religion~\cite{dash2022insights, rifat2024politics}, caste~\cite{vaghela2021birds, vaghela2022caste, panda2020affording}, and socioeconomic status~\cite{kommiya2022voting, seth2022closed}. Platform features, algorithms, metrics, moderation practices, and patterns of participation can shape whose voices are amplified, whose perspectives are marginalized, and how discourse unfolds~\cite{haimson2015disclosure, kumar2018uber, das2021jol, harris2023honestly}. As a result, the same platforms that enable decolonial discourse can also constrain it, producing a dual dynamic of empowerment and limitation~\cite{das2024identity, das2023decolonization}. This paper seeks to understand how that is shaped by audience interaction, visibility, platform governance, and state power.

\subsection{Platform Governance as a Structuring Force on YouTube}
While video-mediated platforms such as YouTube enable new forms of cultural expression and decolonial discourse~\cite{das2024reimagining}, they are also structured by governance mechanisms that shape what content is produced, circulated, and sustained, since these platforms operate as sociotechnical systems that can organize participation through economic incentives, policy frameworks, and technological affordances~\cite{gillespie2010politics, burgess2018youtube}. Therefore, cultural production on YouTube must be understood in relation to the platform's broader political economy and governance structures.

YouTube is emblematic of platform capitalism, where value is generated through data extraction, targeted advertising, and the coordination of interactions among creators, audiences, and advertisers~\cite{srnicek2017platform, zuboff2023age}. Within this system, content creators are not only cultural producers but also participants in a form of platform labor, where visibility, engagement, and monetization are closely intertwined~\cite{ma2021advertiser, simpson2023rethinking}. Consequently, creators often strategize their content in response to advertiser preferences, audience metrics, and algorithmic visibility~\cite{kumar2019algorithmic, joseph2024advertising}.

A key dimension of this governance is monetization, which introduces uneven economic incentives across different audiences and regions~\cite{duffy2017not}. Advertising-based revenue models differentially value audiences according to regional advertising markets, including local purchasing power, advertiser demand, and competition for advertising placements~\cite{joseph2024advertising}. Creator payouts are thus tied to what advertisers can and are willing to pay to reach particular audiences. Lower advertising costs may make the platform more accessible to local businesses while simultaneously generating less revenue for creators whose audiences are concentrated in those regions~\cite{zhang2015analyzing, chen2014economic}. For example, reported YouTube ad-revenue payout per thousand impressions can range from less than a dollar in regions such as Bangladesh (\$0.52), India (\$0.96), and Pakistan (\$0.42) to about \$12 in the United States~\cite{ahmed2025youtube}. While these disparities emerge from an interconnected advertising ecosystem involving platforms, creators, advertisers, audiences, and regional economic conditions, they can influence which audiences creators target and which forms of cultural production are economically sustainable, potentially reinforcing global inequalities in representation and visibility.

In addition to economic incentives, content moderation practices play a central role in structuring participation on YouTube~\cite{childs2023examining}. Platform governance is enacted through formal policies, automated systems, and discretionary enforcement, which together determine what content is permissible, visible, or monetizable~\cite{gillespie2010politics, jiang2023trade}. Prior work has shown that these processes are often opaque and inconsistently applied, making it difficult for creators to anticipate or interpret platform decisions~\cite{fiesler2019creativity, fiesler2023chilling}. This opacity can create uncertainty and constrain creative expression, particularly among creators engaging with sensitive or politically charged topics~\cite{das2021jol, van2023investigating}, often without achieving the stated objective of curbing radical content~\cite{hosseinmardi2021examining, pisharody2025changes}.

Copyright governance further complicates cultural production on YouTube, especially for creators producing transformative or derivative content (e.g., reaction videos)~\cite{vogele2017s}. Legal frameworks such as the Digital Millennium Copyright Act underpin platform mechanisms, including notice-and-takedown systems and automated detection tools (e.g., Content ID), which scan uploaded content against proprietary databases~\cite{uwnddmca, perel2015accountability}. Though these systems are designed to protect intellectual property, creators often struggle to understand and navigate copyright rules, leading to disputes, content removal, and self-suppression~\cite{fiesler2014copyright, fiesler2015understanding, fiesler2020lawful}. These challenges are particularly salient for creators whose work involves remixing, commentary, or critique, where the boundaries of fair use remain ambiguous~\cite{fiesler2023chilling, vogele2017s}. Given these ambiguities, users and creators develop ``folk theories"--informal, experience-based understandings of how platform systems operate--which shape how they navigate visibility, moderation, and participation~\cite{eslami2016first, devito2017algorithms}.

Beyond platform-level policies, YouTube's governance ecosystem includes intermediary actors such as multi-channel networks (MCNs), which introduce additional layers of power and control. MCNs are third-party service providers that affiliate with multiple YouTube channels and offer services such as audience development, content programming, creator collaborations, digital rights management, monetization, and/or sales~\cite{youtubendmulti}. Industry reports suggest that up to 75\% of top-ranking YouTube search results are associated with MCN-affiliated channels~\cite{allndtier}, although these affiliations are not readily visible to viewers. MCNs are not endorsed by YouTube or Google, but YouTube maintains a list of certified ones~\cite{youtubendmulti}. They often leverage their relationships with platforms and advertisers to benefit affiliated creators~\cite{lobato2016cultural, graves2016law, xiao2025institutionalizing}. However, this intermediary structure can also produce asymmetries: larger or better-resourced channels may gain advantages in visibility, revenue, and dispute resolution, while smaller creators face greater barriers to participation~\cite{siciliano2023intermediaries}. Thus, MCNs function not only as support structures but also as gatekeepers that shape the distribution of opportunities and authority within the platform.

Recent research further shows that creators do not simply operate within these systems but actively engage with and reshape them through everyday practices. They also perform ``critical infrastructuring," i.e., modify and extend platform features to address limitations in accessibility and visibility~\cite{simpson2022hey}. In this sense, platform governance operates not only through formal policies or intermediaries, but also through the underlying infrastructural systems that shape and constrain action. As infrastructures often become visible only when they break down or are being actively reconfigured~\cite{star1999ethnography, plantin2018infrastructure}, content creators must navigate and adapt to these in practice~\cite{simpson2022hey}.

Importantly, platform governance does not operate in isolation from broader sociopolitical contexts. Instead, platforms often mediate between users and external regulatory pressures, producing arrangements in which governance is distributed across multiple actors and levels~\cite{gillespie2010politics}. Their regulation of users is embedded within layered systems that include legal frameworks, institutional actors, bureaucratic infrastructures, and state influence~\cite{fasel2025between, gorwa2024politics}. Hence, content production is shaped not only by platform policies but also by intermediary enforcement, the state's regulatory expectations, and political pressures. Together, monetization policies, moderation practices, copyright regimes, and intermediary institutions (e.g., MCNs) do more than constrain or enable content creation. They differentially shape whose voices are amplified, whose content remains sustainable, and whose claims can be defended within the platform ecosystem. However, existing research has paid limited attention to how these layered forms of governance intersect with postcolonial identity tensions and decolonial discourse, particularly in transnational contexts. In this paper, we examine how YouTube creators engage diverse audiences, navigate platform governance, contend with state political influence, and sustain their work toward decolonial discourse amid these challenges.
\section{Methods}
This paper is part of a broader multi-platform research project examining how computing systems (e.g., online communities, video-sharing platforms, and algorithmic infrastructures) shape the expression of identities within colonially marginalized communities~\cite{das2024reimagining, das2023decolonization, das2024identity, das2023studying, das2022collaborative, das2021jol}. While such systems can enable new forms of identity articulation and decolonial discourse, they are also embedded within historical power relations, economic asymmetries, and uneven cultural epistemologies that continue to reflect colonial legacies. In this paper, we focus on how decolonial discourse is sustained on YouTube within the Bengali geocultural context. Specifically, we examine how creators navigate layered sociotechnical constraints arising from audience publics, platform governance, and state-linked intermediaries. We recruited participants from Bangladesh, India, and Pakistan who create YouTube videos related to Bengali identity, culture, and society. These regions provide a critical site of inquiry, as they are shaped by shared colonial histories, post-partition national formations, and ongoing tensions across religious, linguistic, and national identities. 

Given the effectiveness of video-mediated discourse in these contexts~\cite{das2024reimagining}, we focus on YouTube as a key site for examining how such discourse is produced, contested, and sustained. YouTube's widespread adoption~\cite{globalmediainsight2023YouTubeStatistics}, accessibility to both amateur and professional creators~\cite{yuan2022what}, and role in facilitating sociopolitical conversations among marginalized communities~\cite{das2024reimagining} make it an appropriate platform for this study. Through a qualitative study based on semi-structured interviews, we examine how creators encounter and negotiate social, economic, and political constraints in sustaining decolonial discourse. Prior to conducting this study, we received approval from our university's institutional review board for all materials and procedures.

\subsection{Recruitment}
Data for this study comes from semi-structured interviews with 15 YouTube content creators residing in Bangladesh, India, or Pakistan. Our eligibility criteria required that participants (1) be 18 years or older, (2) actively create YouTube content related to Bangladesh, India, and/or Pakistan, and (3) reside in one of these countries. Participants' channels did not fit a single genre. Across the sample, creators produced political explainers, journalistic reports, satire, travel videos, social interviews, public reaction videos, and reaction videos to existing media. The topics they addressed included Bengali culture and language, India--Bangladesh--Pakistan relations, religious and national identity, social experience, political issues and events, historical events, popular media, festivals, food, and everyday cultural practices. We treated these videos as relevant to decolonial discourse not because of their genre, but because participants used them to engage questions of cultural belonging, historical memory, postcolonial division, and representation across Bengali communities. For example, in our prior work~\cite{das2024reimagining}, we found that reaction videos can serve as a form of decolonial discourse when creators used existing media artifacts to interpret shared histories, compare cultural practices, or build understanding across audiences in Bangladesh, India, and Pakistan.

The first author, who is from Bangladesh and an active YouTube user, brought contextual familiarity to the recruitment and interview process. We identified participants using a combination of purposive sampling~\cite{suri2011purposeful} and snowball sampling~\cite{goodman1961snowball}. Prior works have highlighted that different linguistic, religious, and national identities are central to online decolonial discourse within the Bengali geocultural context~\cite{das2021jol, das2022collaborative}. Therefore, we began by searching YouTube using combinations of Bengali identity-related keywords such as \textit{Bengali}, \textit{Bangladesh/Bangladeshi}, \textit{India/Indian}, and \textit{Pakistan/Pakistani}, following approaches used in prior work~\cite{das2021jol}, in both Bengali and English. Although many of the retrieved results were Bengali movies, dramas, and music, since our focus was on discourse, we identified channels that produced commentary-, discussion-, and analysis-oriented content rather than entertainment media such as films or music. We also leveraged YouTube's recommendation system to identify related channels and expand the diversity of our sample following suggestions from prior work~\cite{das2022note}. Additionally, we recruited participants through personal networks and participant referrals.

To contact participants, we collected publicly available contact information from YouTube channel descriptions, including email addresses and social media handles (e.g., Facebook, Instagram, Twitter). We initially reached out through email with a recruitment flyer describing the study, and followed up after one week if needed. When email contact was unsuccessful, we reached out using social media platforms. Following each interview, we asked participants to recommend additional creators, enabling snowball sampling. Recruitment continued until we reached theoretical saturation, resulting in total 15 participants. Their demographic details are summarized in Table~\ref{tab:demography}.

\begin{table}[!ht]
    \centering
    \caption{Demographic information of the participants}
    \label{tab:demography}
    \begin{tabular}{p{1.4cm}p{1cm}p{1.8cm}p{1cm}p{2.15cm}p{1.7cm}p{2.3cm}}
    \toprule
        \textbf{Identifier} & \textbf{Gender} & \textbf{Country of nationality} & \textbf{Age} & \textbf{Religion} & \textbf{Education} & \textbf{Occupation} \\
    \midrule
        P1 & Male & India & 20-24 & Muslim & Bachelor's & Engineer \\
        P2 & Male & India & 35-40 & Hindu & Master's & Journalist \\
        P3 & Male & Pakistan & 25-30 & Muslim & Bachelor's & Student \\
        P4 & Male & Bangladesh & 25-30 & Muslim & Master's & Journalist \\
        P5 & Male & Bangladesh & 30-34 & Muslim & High school & Freelancer \\
        P6 & Female & India & 30-34 & \textit{Did not disclose} & Bachelor's & TV presenter \\
        P7 & Male & Bangladesh & 30-34 & Muslim & Bachelor's & YouTuber \\
        P8 & Male & Bangladesh & 30-34 & Muslim & Master's & Journalist \\
        P9 & Male & India & 40-44 & Hindu & Master's & Govt. employee\\
        P10 & Male & Pakistan & 25-30 & Muslim & Bachelor's & Engineer \\
        P11 & Female & Bangladesh & 25-30 & Muslim & Master's & Job Aspirant \\
        P12 & Female & India & 20-24 & Hindu & Bachelor's & Student \\
        P13 & Male & Pakistan & 30-34 & Muslim & Master's & Engineer \\
        P14 & Male & India & 20-24 & Hindu & Master's & Web-developer\\
        P15 & Female & India & 20-24 & Hindu & Master's & Student\\
    \bottomrule
    \end{tabular}
\end{table}

\subsection{Interviews}
We conducted 15 in-depth semi-structured interviews between Summer 2020 and Summer 2022, following qualitative methodologies outlined by Strauss and Corbin~\cite{strauss1994grounded} and Yin~\cite{yin2017case}. Interviews were designed as life histories~\cite{wengraf2001qualitative}, situating participants' content creation practices within their broader lived experiences. We began with demographic questions, followed by questions about participants' trajectories into content creation and their experiences engaging with audiences and platforms. We then focused on how participants navigate audience interactions, platform policies (e.g., monetization, copyright, moderation), and the broader sociopolitical contexts that shape their work. This included probing how they manage challenges, maintain presence, and sustain engagement.

Given the geographic distribution of participants, interviews were conducted via Zoom or telephone, depending on participants' preferences and connectivity conditions. Participation was voluntary, and participants were not compensated. Prior to each interview, participants provided verbal consent and agreed to be audio-recorded. Interviews were conducted in Bengali, English, or Hindi/Urdu\footnote{As spoken languages, Hindi and Urdu are mutually intelligible, to the point that they are sometimes considered dialects or registers of a single spoken language known as Hindi-Urdu or Hindustani~\cite{editors2022hindustani, wiki2022hindi}.} based on participant preference. Interviews lasted between 30 minutes and 1 hour 53 minutes, with an average duration of approximately 60 minutes. The first author, a native Bengali speaker with bilingual proficiency in English and working proficiency in Hindi/Urdu, conducted all interviews. All interviews were transcribed and, where necessary, translated into English. The first author translated Bengali interviews, while the Hindi/Urdu interviews were translated by a native Hindi speaker. All transcripts were anonymized and de-identified prior to analysis.

\subsection{Data Analysis}
We analyzed the data using an inductive, grounded theory-inspired approach~\cite{strauss1994grounded}, commonly used in HCI and social computing research~\cite{das2026global, houston2016values, das2021jol}. Following Strauss and Corbin's three-phase approach~\cite{strauss1994grounded}, we conducted open, axial, and selective coding.

In the open coding phase, the first author iteratively reviewed transcripts to identify recurring concepts. These open codes were often related to participants' experiences with audiences, platforms, and institutional structures. Examples of open codes included \textit{``audience distrust because of YouTuber's demographic identity"}, \textit{``accusation of hidden motive"}, \textit{``harassment and safety concerns"}, \textit{``monetization inequality"}, and \textit{``platform opacity and uncertainty"}. Both authors met weekly to discuss and refine emerging codes. During axial coding, we grouped related open codes into higher-level categories. For example, we merged the first two open codes mentioned earlier under \textit{``tension with the audience"}. In the selective coding phase, we examined relationships among these axial codes to develop themes that capture how layered sociotechnical constraints shape the viability of decolonial discourse and how creators negotiate the constraints and challenges related to audience, platform, and state in practice.

Since interview data is contextual, consistent with interpretivist qualitative traditions, we did not calculate inter-coder reliability~\cite{mcdonald2019reliability, sebastian2019distinguishing}. Our analysis is reflexive and acknowledges that interpretations are shaped by researchers' positionalities and scholarly orientations (see Section~\ref{sec:positionality} for details about authors' backgrounds)~\cite{anderson1994representations}. Thus, our findings reflect both participants' experiences and our interpretive lens grounded in postcolonial and decolonial computing perspectives.

\subsection{Positionality and Reflexivity Statement}\label{sec:positionality}
When researching marginalized communities, the authors' racial and ethnic backgrounds may influence their perspectives and interpretation~\cite{schlesinger2017intersectional, liang2021embracing}. The first author is a cisgender, heterosexual man from the Bengali Hindu minority community in Bangladesh, with a family history affected by refugee crises because of the partition of the Indian subcontinent in 1947 and the liberation war of Bangladesh in 1971. In addition to designing the study, he interviewed participants and led the data analysis. The second author is an Iraqi-American cisgender heterosexual man from a minority group within Iraq who contributed to the study as the anchor author. He was deeply involved throughout the study from its initial inception and design through the writing of this manuscript. Given their embeddedness in colonially shaped sociocultural contexts that motivated their research, the authors acknowledge that their work, which focuses on conversations about colonial histories, is inherently political. Both authors' memberships in various minority communities, lived experiences in colonially impacted societies, and prior experience in critical (e.g., decolonial, postcolonial) computing research motivated their study of colonially marginalized communities' practices with technology. Furthermore, the authors acknowledge that the University of Colorado Boulder, where this research was conducted, is situated on the land of the indigenous Arapaho, Cheyenne, and Ute peoples.
\section{Results}
We examined how Bengali YouTubers producing sociopolitical content in postcolonial contexts navigate a layered sociotechnical environment shaped by transnational audiences, platform governance, and state-linked regulation. Participants' accounts reveal their relational labor to cultivate and maintain audience ties, boundary work to negotiate identity-based expectations, and infrastructural work to sustain circulation under unequal platform economies and political risk.

\subsection{Audience Publics as Sites of Postcolonial Tension}
Our participants discussed the challenges stemming from their identities that affect their relationships with audiences, as well as their strategies for audience management and interaction. We examine these dynamics through the lens of postcolonial tension, which refers to the conflict and struggles for identity, power, and cultural recognition that arise in societies historically shaped by colonial rule and values. Across our interviews, audience publics emerged not simply as passive consumers of content, but as active sites where such tensions are expressed, negotiated, and enforced.

\subsubsection{Challenges from Fragmented Publics and Legitimacy Policing}
First, we discuss various ways our participants experience challenges in building relationships and interacting with their audience. Tensions among various identities across dimensions such as gender, religion, nationality, language, and cultural practices are a byproduct of this region's colonial past. Because of how colonization has historically shaped and fragmented Bengali societies, YouTubers face challenges in representing the pluralism and intersectionality of Bengali culture.

\noindent\paragraph{Distrust among Religions}
Religion has historically served as a primary axis of division in the region, fostering heightened distrust among communities~\cite{chatterjee1993nation, rifat2024politics}. Similar distrust and fragmentation mediate YouTubers' interactions with their audiences. For example, P1 is a Muslim YouTuber from India who makes videos about the positive aspects of Bangladeshi societies and their recent development. However, some of his audiences from his own country have alleged that his videos promote a Muslim-majority country because of his religious identity. Instead of appreciating the YouTubers' efforts, audiences often view their topics and motivations through a lens shaped by postcolonial tension among religious communities.

Similarly, some of the audiences of Pakistani participants, P3, P10, and P13, abused them for focusing on improving relationships with Bangladesh and India. Their audiences did not appreciate that these YouTubers, being from Muslim-majority Pakistan, wanted to improve relationships with Hindu-majority India. These reactions illustrate how historical and geopolitical tensions continue to shape expectations about who can speak about whom and in what ways.

\noindent\paragraph{Linguistic and Nationalistic Identities}
Tensions around legitimacy also emerge through linguistic and national identities. For example, P2, a Bengali Hindu YouTuber from India, creates videos critiquing contemporary Indian politics. He explained that some viewers responded to these critiques by questioning whether, as a Bengali, he could legitimately convey Indian perspectives. P2 contrasted how nationalist audiences displace critics according to their perceived linguistic and religious identities:

\begin{quote}
    According to them, Bengalis are not Indian enough, so I am not Indian enough. They'll say, ``Oh, you are a [common Bengali Hindu surname]!" So, there are two qualifiers [religion and language]. To them, Bengali equals Bangladeshi. ... [Compared to someone speaking a Hindustani language, who critiques state policies], I am not sent to Pakistan; by the way, I am sent to Bangladesh. So, ``not enough" identity is something that I completely resonate with. It is all about identity--``my identity is more than your identity." \hfill (P2, male, India)
\end{quote}

Unlike the cases we discussed before, here, P2's videos are not questioned through the lens of religion. Although his Hindu surname made one aspect of his identity legible within Hindu-majority India, viewers' conflation of Bengali identity with Bangladesh positioned him as insufficiently Indian. As nationalist audiences symbolically assign dissenters to different neighboring countries according to their perceived identities, the issue was not simply interpersonal abuse but legitimacy policing. P2's Bengali identity became grounds for questioning his national belonging and, consequently, his authority to criticize Indian politics. Such frictions within collective identities become particularly complex given the transnational character of most participants' audiences. Most participants reported that their audiences are primarily from Bangladesh, followed by India and Pakistan, with sizeable additional viewership from diaspora communities in the Gulf, the UK, and the USA. When our participants' discourses address broader regional geopolitical issues, parts of their audience often question their knowledge or intentions, driven by confirmation bias and nationalist sentiment.

For example, P9, a history-focused YouTuber, makes videos about the precolonial, colonial, and postcolonial history of Bengal. He talked about how audiences challenge historical narratives that do not align with their prior beliefs, a pattern often established by fragmented narratives in history curricula across different countries~\cite{das2024reimagining}. He said:

\begin{quote}
    They would say that I was spreading misinformation. Then, I asked them to give the correct information. But, they could not respond. There are also times when something is in history, but they are not willing to accept it. They hold on to their personal, biased beliefs, and they won't come out of it. \hfill (P9, male, India)
\end{quote}

In some cases, these tensions escalate into coordinated forms of backlash. Participants described instances where geopolitical events triggered collective audience actions. For example, P13 shared:

\begin{quote}
    When the standoff with India happened, there was a campaign of Indians unsubscribing from Pakistani channels, and we were victims of that as well. We lost our 4000 subscribers at that moment. \hfill (P13, male, Pakistan)
\end{quote}

Such incidents illustrate how audiences mobilize along postcolonial national identities to regulate discourse and sanction content creators. Hence, YouTubers consequently calibrated future content around not only anticipated popularity, but also the possibility that cross-border or counter-national narratives would be interpreted as national disloyalty, foreign allegiance, or insufficient belonging.

\noindent\paragraph{Gendered and Cultural Policing}
Postcolonial tensions also intersect with gender and class, shaping how creators are perceived and treated by audiences. Participants described frequent ad hominem attacks targeting appearance, accent, and dress. Female creators, in particular, faced gendered harassment and safety constraints. For example, P15 talked about several instances of her Bangladeshi male audiences categorizing her sleeveless dress as short and abusing her with vulgar comments. All four of our female participants talked about the gendered aspect of the harassment, obstruction, and fear they face in their content creation. Participant P11 explained how such norms restrict her ability to conduct fieldwork:

\begin{quote}
    Many people would certainly obstruct me. They will scold me, saying, ``You are a woman, and you came to make videos!" ... I cannot go to collect footage for those videos at night. \hfill (P11, female, Bangladesh)
\end{quote}

Participants also encountered classed and cultural hierarchies in audience responses. For example, P14 described how viewers dismissed certain cultural figures as representing ``uncultured village people," which reinforces the relegation of rural cultural practices in colonial Bengal:

\begin{quote}
    Some rebuked us for highlighting [a popular Bangladeshi media personality]'s work. They told us that people in cities do not follow her work and that they are only for the uncultured village people. \hfill (P14, male, India)
\end{quote}

These reactions reflect historically rooted distinctions between elite and subaltern cultural forms, shaping which representations are considered legitimate.

These accounts do more than show that audiences influence creators' content decisions. Here, legitimacy depends partly on who is speaking: an Indian Bengali critic can be positioned as Bangladeshi, Pakistani reconciliation content can be interpreted as nationally disloyal, women's presence in public space can be treated as problematic, and rural cultural representation can be dismissed as insufficiently respectable. Hence, YouTubers calibrated future content not merely around popularity, but around whether their narratives would be interpreted as foreign allegiance, insufficient belonging, or illegitimate cultural representation.

\subsubsection{Relational and Discursive Strategies for Sustaining Audience Publics}
In response to these challenges, our participants engaged in relational labor with discursive boundary work. We use relational labor narrowly to describe efforts to cultivate and maintain ongoing audience relationships, including culturally situated greetings, linguistic accommodation, responsiveness to requests, and community interaction. We distinguish these practices from boundary work through which participants anticipated identity-based judgments, decided which topics to address, and managed the risks of being interpreted across conflicting national and religious publics.

\noindent\paragraph{Finding Support from Community}
Despite challenges, participants described strong support from their audiences, families, and social networks. This support provides emotional encouragement and, in some cases, material and logistical assistance. For example, P12 described how interactions with viewers foster a sense of belonging:

\begin{quote}
    Some say, ``Sister, if you come to Bangladesh, come and stay at our home. We will take you to visit different places in Bangladesh." ... It feels like they are my own family members. \hfill (P12, female, India)
\end{quote}

In addition to online encouragement, participants facing safety risks, such as female creators, trusted people in their existing social networks. Besides those who appreciate their content online, they often offer help. For example, P11 described relying on alumni and acquaintances for protection and logistical support:

\begin{quote}
    There are many alumni from our university in high-level administration. They can be high-ranked bureaucrats or police officers, but for us, they are like our brothers and sisters. ... We consider whether we can go there safely. \hfill (P11, female, Bangladesh)
\end{quote}

\noindent\paragraph{Following Identity Norms and Customs}
Participants also adapt their self-presentation to align with audience expectations across religious, linguistic, and national identities. This includes using culturally appropriate greetings, languages, and accents to establish affinity. For example, given the variation and strong association of salutations with different religious communities, P3 uses common phrases for Bengali communities and specific ones for Muslims (e.g., \textit{Assalamualaikum}) and Hindus (e.g., \textit{Namaskar}). Some of our participants also start the videos with the primary language, dialect, and accent (e.g., \textit{Bangal} or \textit{Ghoti}) of their audiences. For example, P10 described how using Bengali helps build a connection:

\begin{quote}
    Sometimes, I try Bengali in the beginning. It attracts the viewers and makes them feel connected to our channel. ... It's a kind of loving gesture from us. \hfill (P10, male, Pakistan)
\end{quote}

However, since some of our participants (e.g., those from Pakistan) do not speak Bengali as their native tongue, or because the accent of their audience (e.g., Indian Bengali participants who do not speak the Bangladeshi form of the language) differs, they also try to be mindful not to mispronounce the language. Some of them also take help from their friends who speak the languages natively in this task. Participants also provide subtitles, especially when addressing transnational audiences.

\noindent\paragraph{Deciding on the Dilemma of Content Topics}
To mitigate identity-based tensions, participants strategically decide what topics to engage with. Some participants passionate about the decolonial aspect of their YouTube video-making develop a range of creative strategies. Considering ideological stubbornness as one of the main impediments to political discussions, participant P2 created a recurring caricature of a strongly partisan supporter, who depicts a hyperbolic figure holding uncritical loyalty to a political ideology. Through that character, he speculated, explored, and demonstrated how his videos could be interpreted in different political echo chambers. For example, P2 described how that helps him deflect ideological backlash:

\begin{quote}
    I used to think about how those could be taken out of context... what logic someone like [the caricature] would apply to this. Of course, my school and college acquaintances who share characteristics of that caricature in the WhatsApp group gave me enough understanding of how they think. ... That is how the whole concept came around. \hfill (P2, male, India)
\end{quote}

Others (e.g., P13, P14, P15) avoid political or religious discussions altogether or delegate moderation to trusted audience members known through other platforms (We will discuss the multi-platform practice in the next section). While these strategies reduce conflict and broaden appeal, they also constrain engagement with issues central to decolonial discourse.

Overall, these practices highlight how creators perform continuous boundary work to sustain their presence on the platform. Rather than operating outside audience dynamics, they actively negotiate identity, legitimacy, and risk in response to them.

\subsection{Platform Governance as Postcolonial Structure}
Our participants discussed various challenges that emerge from the mediation of postcolonial economic and institutional structures through YouTube. Their accounts show how geographically differentiated platform economies and institutional hierarchies condition whether their efforts to represent marginalized cultural forms and sustain cross-border Bengali discourse are economically viable and defensible. They also explained how they overcome, overlook, and navigate through these challenges by strategizing their use and presence on online platforms. Overall, in this section, we examine platform governance not as a direct reaction to decolonial discourse, but as an unequal infrastructure through which that discourse need to circulate.

\subsubsection{Challenges from Differential Policies and Institutional Asymmetries}
YouTube's policies determine how YouTubers' content is moderated, how they are compensated, and who gets recognized as a content creator. We examine how these dynamics constrain their expressive choices and can reinforce existing geographic, economic, and institutional hierarchies within Bengali cultural production.

\noindent\paragraph{Unequal Monetization System}
Our participants discussed how YouTube's monetization policy is discriminatory and provides significantly different financial advantages to them based on the location of their viewers. They explained that because more companies in the Global North advertise on YouTube than in the Global South, ad revenue in the Global South is lower than in the Global North. Since one of the motivations for YouTubers is to earn income from their content, platform capitalism strongly influences their decisions about whom they make content for. For example, within Bengali communities, our participants prioritize making videos that interest the Bengali diaspora over local Bengali communities in Bangladesh or India. P14 explained:

\begin{quote}
    If someone watches a video from the US or the UK, our earnings will be higher than those of someone watching it from Bangladesh. The ads on a video depend on the country [where one is watching from]. The earnings will depend on ad prices in that country. \hfill (P14, male, India)
\end{quote}


Since YouTubers prioritize the cultural values and practices of the Bengali diaspora living in the Global North, this normalizes the representation of a particular set of Bengali cultural preferences, practices, and identities online. The diasporas in these locations represent an archaic or selective set of Bengali cultural preferences, which does not portray the shifts in local Bengali culture in Bangladesh and India. This also exacerbates the marginalization of some Bengali communities, such as rural and agrarian communities, to a subaltern space. While the representation of diasporic practices is important, in prioritizing the diaspora for the sake of higher financial incentives, the YouTubers commodify their viewers from a commercial perspective. Consequently, the platform's differential monetization policies shape what our participants represent in their discourse on Bengali sociocultural topics and issues.

\noindent\paragraph{Inconsistent and Opaque Content Moderation}
Contrary to the rationales about how YouTube monetizes videos, our participants were often unclear about many other platform policies. They were often frustrated by the opacity of content moderation, such as deletions or geographic blocks. For example, P14, a participant from India who primarily makes videos about the transnational aspects of Bengali culture, wondered why some of his videos are blocked in Bangladesh while viewers elsewhere can see them. Such uncertainty around whether the audience for whom, or based on whose requests, the YouTubers made certain videos would be able to see them poses a challenge for creators. They often emailed YouTube to ask for clarification about its decision regarding their content. However, they often found YouTube's response to be delayed and unhelpful.

In addition, YouTube's moderation process is often hands-off, uncertain, and inconsistent, especially regarding copyright claims. Because some of our participants make reaction videos that reuse parts of previously published content, they are particularly prone to copyright claims. Our participants shared that while some channels are lenient toward such content creators, others are more aggressive about claiming copyright. They also described YouTube as taking a hands-off approach to settling copyright claims, often suggesting that content creators resolve disputes among themselves while the platform abides by their decisions.



\noindent\paragraph{MCNs' Asymmetric Power}
Our participants highlighted how institutions such as MCNs affect their work. In their opinion, usually more reputable, larger, and more financially solvent YouTube channels join MCNs. They described incidents in which an MCN-member YouTube channel used footage from their videos and later claimed copyright over those same videos. Participant P4 critiques YouTube's copyright policy as ``weak" and describes how the prioritization of the claims from MCN-member channels discriminates against them:

\begin{quote}
    YouTube blindly trusts one who has MCN. ... YouTube believes that a channel has MCN, so whatever that channel uses belongs to [that channel]. ... You can sometimes be blamed under copyright policy for using your own content. \hfill (P4, male, Bangladesh)
\end{quote}

Based on such incidents, our participants found YouTube to be complicit in reflecting economic hierarchies onto Bengali culture. As frustrated participants perceived a lack of responsibility from YouTube for mediating copyright claims, they viewed YouTube as technologically complicit in hierarchical cultural logic. Especially given the power hierarchy among YouTube channels affiliated with MCNs, our participants proposed that YouTube play a more active role in these matters and leverage video metadata to arbitrate disputes. P11 shared her concern and proposal as follows:

\begin{quote}
    Do you think the other party who stole my video will tell YouTube that they stole my content and they are the ones at fault? ... YouTube can easily check who uploaded the content first, but they do not do it. I uploaded a video first, then someone else downloaded it and uploaded it. They make me the party at fault with their power of MCN. \hfill (P11, female, Bangladesh)
\end{quote}

Taken together, these findings show that platform governance operates not only through differential monetization and opaque moderation but also through institutional asymmetries that make some creators' claims more credible and enforceable than others. In that sense, YouTube does not simply mediate decolonial discourse; it stratifies whose discourse can be sustained, defended, and monetized.

\subsubsection{Strategies for Navigating Platform Economies and Infrastructures}
Our participants negotiate among the demands of their audience, the potential for monetization, and the possibility that their videos will be blocked or claimed for copyright. They often strategize their video-making by prioritizing non-monetary incentives, acting based on collective folk theories, and utilizing an ecosystem of multiple platforms.

\noindent\paragraph{Prioritizing Non-monetary Incentives}
Because of YouTube's monetization policy, our participants were likely to earn more by making videos appealing to viewers in the Global North. Considering the financial potential, making videos for viewers in the Global South locations is not beneficial for them. However, most of their audiences come from countries like Bangladesh and India. Therefore, although ad revenue per view from these countries is low, making videos that interest viewers there often leads to more engagement, such as likes, comments, and shares. Therefore, they often shift from focusing on the monetary gain from a particular video to using it as an opportunity to invest in their channels' future growth and reputation.

Reaction videos are quite popular among our participants' Bangladeshi and Indian audiences. Hence, despite these videos' viewers being at low-revenue locations and the possibility of copyright claims by MCNs, our participants believe that making these reaction videos demanded by their viewers helps them ``build affinity" with their audiences. P15 explained her rationale:

\begin{quote}
    In some cases, so many viewers request that we make a reaction video that we must make one. All our hard work behind that video gets lost because those videos are not monetized. Even if we sacrifice the money, it will attract many people and help us gain more subscribers. Therefore, although we are likely to get copyright claims, we make some videos to gain subscribers. \hfill (P15, female, India)
\end{quote}

By conceptualizing a video's success in terms of non-monetary incentives, our participants prioritize increasing subscribers and engagement over immediate revenue. Using this strategy, YouTubers overlook some of the monetization challenges stemming from the platform's policies.

\noindent\paragraph{Acting based on Collective Folk Theories}
Due to YouTube's lack of transparency in its policies, the participants discussed their experiences of being monetarily penalized or having content blocked by other content creators. Through this, they developed collective folk theories about the reasons for certain behaviors or decisions by YouTube's algorithms and policies. For example, P4 explained one of his team's past experiences. They made a video report about an allegedly corrupt local officer. To protect the privacy of a minor victim, they blurred the child's face in that video. When that video was ``yellow monetized"\footnote{A yellow icon in YouTube Studio indicates limited or no ad revenue for content not meeting advertiser-friendly guidelines.}, they wanted to understand the reason. He said:

\begin{quote}
    We blurred a kid's face because we did not want to show them. ... [Someone I consulted with] told me that it happened because I have a blurred face in my video and suggested removing the blurred face. ... We did not know YouTube would do so. It is not mentioned anywhere in their policy. Some things with YouTube's policy are eccentric. They do whatever they want. \hfill (P4, male, Bangladesh)
\end{quote}

Therefore, participants drew on collectively developed understandings of platform enforcement to reduce the likelihood of copyright claims. Alongside standard practices such as crediting original creators, they adopted additional tactics that they believed might affect automated detection, including making reused footage semi-transparent or lowering its resolution. These practices reflect participants' attempts to manage uncertainty around copyright enforcement, while highlighting their own contributions through opinions and reactions.

\noindent\paragraph{Utilizing a Combination of Multiple Platforms}
Given the understanding that online platforms are subject to different regulatory institutions and policies, a common strategy among our participants was to build a multi-platform ecosystem to disseminate their videos. For example, they often share videos on Facebook, which they believe ``is much more easygoing in terms of copyright than YouTube" (P14, male, India). Such a multi-platform approach also helps YouTubers reach viewers of diverse demographic backgrounds. As the same content is regurgitated across multiple platforms, such as YouTube and Facebook, it provides YouTubers with multiple revenue streams, helping them overcome the challenges of lower monetization for Global South-related content. While some participants use multiple platforms to post identical content, some strategize by using different platforms for different purposes, such as Facebook groups to build community with their audiences, Facebook pages to share promos of upcoming YouTube videos, Twitter to test the waters on contemporary events and post one-line punches, and Instagram for fun content like memes.

\noindent\paragraph{Seeking Alternative Revenue Streams}
Some of our participants also set up online financial platforms, such as Patreon, Paytm, and PayPal, associated with their YouTube channels. Different strategies, such as live sessions and merchandise, have provided better financial support for some participants. Besides overcoming financial challenges, these premium channels of interaction also helped them avoid some of the abuses stemming from postcolonial tensions. P2 explains this dual benefit as follows:

\begin{quote}
    Live sessions act as a revenue source because I answer questions that are paid for. What I have discovered is that people don't pay you money to abuse you, so I get a good bunch of questions from people. \hfill (P2, male, India)
\end{quote}

YouTube did not targeted our participants because their content was decolonial, but our findings illustrate how their cultural production took place within a platform economy that assigned lower advertising value to local audiences, rewarded attention from comparatively privileged diasporic audiences, and gave institutionally connected channels greater capacity to enforce ownership claims. For participants seeking to represent cross-border, rural, or otherwise marginalized Bengali experiences, these arrangements affected which audiences were financially valuable and which cultural materials could be reused or defended--overall, which forms of discourse could be sustained.

\subsection{State-Linked Regulation and Political Risk}
Beyond audiences and platform governance, our participants described how state-linked regulatory environments shape their ability to produce and sustain decolonial discourse. These dynamics are not always enacted through direct platform intervention, but often operate through intermediary institutions, informal pressures, and broader political climates. As a result, content creation becomes entangled with concerns about surveillance, registration, access, and personal safety.

\subsubsection{Challenges from Political Constraints and Intermediary Regulation}
Participants discussed how expressions of identity and opinion, even in seemingly trivial contexts, can become subject to policing and legal scrutiny. 

\noindent\paragraph{Anticipatory Fear and Legal Risk}
Participants particularly described how broader political environments shape their perception of risk through both anticipatory fear and the possibility of punitive legal action. Rather than requiring direct intervention, the mere prospect of state attention influences how creators calibrate their engagement with sociopolitical discourse. For instance, P2 emphasized a deliberate effort to avoid visibility to governmental actors:

\begin{quote}
    We don't fancy that our videos are going to reverberate ... The only thing we wish is that it doesn't go to the government, so that we can live in freedom. If it goes to the government, then we are screwed. ... We are afraid of UAPA\footnote{Unlawful Activities (Prevention) Act (UAPA) is India's primary anti-terror law designed to prevent unlawful associations and activities threatening national sovereignty~\cite{ohchrnduapa}. Its amendment in 2019 allows the government to designate individuals as terrorists, not just organizations. It is known for stringent bail conditions, long detention periods without trial, and frequent use against activists and journalists.} being applied, which is the most draconian anti-terror law, and of people being reported to the police. It shows just how bad the situation has become now. \hfill (P2, male, India)
\end{quote}

P2's desired audience was therefore accompanied by an unwanted audience: the state. While several other participants discussed such governmental and political influences, P2 brought them up more frequently, discussed them more, and elaborated on them in greater detail. Among other participants, P8 and P11 mentioned that their family members frequently discouraged them from producing videos engaging with sociopolitical issues, due to concerns about potential negative repercussions from the government and other political parties. For P2, a video could successfully reach viewers yet become dangerous if that visibility extended to governmental actors. This reverses the conventional creator imperative to maximize reach. The problem was not visibility in itself, but the possibility that political critique by a Bengali creator already positioned as insufficiently national could be reclassified as a threat to national sovereignty.

\noindent\paragraph{Governmental Control through Intermediaries}
Although none of our participants had been legally charged prior to their participation in our study, they highlighted that state influence is often mediated through institutional intermediaries rather than direct platform enforcement. In particular, MCNs and other formal or semi-formal entities were described as points through which regulatory expectations and pressures are communicated. For example, P2 described how content creators are often required to register or operate through such intermediaries, which in turn become accountable to governmental authorities:

\begin{quote}
    If you want to scale up, you need to register yourself. ... And the government does not talk to you directly, they talk to the MCNs. So, MCNs become the body that controls everything. \hfill (P2, male, India)
\end{quote}

In this way, MCNs do not function merely as platform-level organizational structures but also as conduits for state oversight. Participants perceived that these arrangements introduce additional layers of control over content production, shaping what can be said and how openly creators can engage with politically sensitive topics. Therefore, they described broader political environments as constraining their work. 

\noindent\paragraph{Legal Risks' Effects on Collaboration}
In contexts where sociopolitical discourse is contentious, creators expressed concern about surveillance, backlash, and potential repercussions. These concerns extend beyond individual creators to their collaborators and networks. For instance, P2 explained the difficulty of recruiting skilled professionals, like graphics designers and video editors, due to the politically sensitive nature of their content:

\begin{quote}
    It is very difficult to find people who are willing to come on camera and talk about these things. People are scared. They don't want to be associated with anything political. \hfill (P2, male, India)
\end{quote}

Such constraints limit not only the range of topics that can be addressed but also the forms of storytelling and collective capacity that are possible. Participants noted that accessing certain locations, institutions, or individuals for content production can be difficult or risky, especially when topics involve a critique of authority or exposure of wrongdoing. Fear reduced who would appear on camera, edit politically sensitive footage, or become publicly associated with a creator. In some cases, creators resort to indirect methods or avoid such topics altogether due to these risks. Thus, decolonial expression was consequently constrained within production networks before a video ever reached platform moderation.

Taken together, these accounts illustrate a spectrum of state-linked pressure, ranging from anticipatory self-censorship to the threat of severe legal consequences. In such contexts, concerns about visibility are not abstract but grounded in the possibility of a disproportionate state response, in which even symbolic or expressive acts may be interpreted as political transgressions.

\subsubsection{Strategies for Managing Risk and Sustaining Participation}
In response to these constraints, our participants developed strategies to manage political risk while continuing to create content. These strategies often involve careful calibration of visibility, collaboration, and distribution. One common approach is to limit direct exposure by controlling how sensitive topics are presented or who is involved in their production. As noted earlier, some participants avoid explicitly political framing or rely on indirect narrative techniques. Others selectively collaborate with trusted individuals or draw on existing social networks to mitigate risk.

Participants also emphasized the importance of infrastructural redundancy as a form of protection. Our participants also have a dominant perception that various online platforms comply with and enforce certain rules differently. If their videos are deleted from YouTube due to government decisions, the videos shared on Facebook help them prove to their audience that they indeed made the requested videos. This also ensures that YouTubers' efforts do not go in vain, irrespective of how the platforms comply with state requests. P2 explains his strategy to resist technologically mediated political influence through a multi-platform presence:

\begin{quote}
    [YouTube] can de-platform you at any given time without warning. I mean, today, the government can write one letter and say that YouTube shut this [channel] down, and they will shut it down. So it is never a good strategy to be on a single platform. Each video we make on YouTube, we upload it to Facebook the next day. \hfill (P2, male, India)
\end{quote}

This strategy is not only about audience reach but also about resilience against potential state-linked interventions. These strategies reflect an ongoing effort to balance visibility and safety. Rather than withdrawing from public discourse, participants adapt their practices to sustain participation under conditions of uncertainty and constraint.

Overall, state-linked regulation shapes not only what creators can say but also how they organize their labor, collaborations, and infrastructures. In this sense, decolonial discourse is not only a matter of representation but also of navigating layered regimes of governance that extend beyond the platform itself.
\section{Discussion}
Audience boundary policing, platform governance, and state-linked political risk are not merely three parallel sources of constraint on Bengali YouTubers but rather form a recursive arrangement. These affect creators' capacity for engagement, assign different economic values to those audiences and different levels of legitimacy to content creators' ownership claims, and turn visibility itself into a source of political danger. Creators respond by adjusting their speech, self-presentation, collaborations, and distribution infrastructures. These adaptations, in turn, shape which narratives audiences encounter and which forms of content remain viable. We describe this condition as the \emph{selective sustainability} of decolonial discourse. That means decolonial expression is not simply enabled or suppressed, but some forms become more sustainable than others because they are easier to make socially acceptable across divided publics, economically viable within platform markets, and politically survivable under conditions of uncertainty. This shifts the analytical question away from whether platforms merely permit marginalized communities to speak--and from uncritically celebrating creators' strategies as resilience--toward examining which forms of discourse can be repeatedly produced, circulated, and defended, and how the strategies that enable participation may simultaneously constrain its political possibilities.

\subsection{Relational Labor under Postcolonial Tension}
Content creation on social media is often conceptualized as a form of relational labor, in which creators engage in ongoing interactions with audiences to build trust, maintain visibility, and sustain engagement~\cite{baym2015connect, duffy2017not, abidin2016visibility}. Our findings extend this perspective by showing that, in postcolonial contexts, relational labor involves negotiating historically produced boundaries around who may represent a culture, narrate a history, or criticize a nation. Attending to these dynamics requires historicism~\cite{soden2021time}: the publics creators encounter are not merely contemporary audience segments but formations shaped by postcolonial partition, nation-building, religious division, linguistic hierarchy, and migration. Hence, the relational labor also becomes historical boundary work.

Participants used Bengali phrases, religiously situated greetings, subtitles, satire, and culturally recognizable forms of address to establish affinities and signal care toward a public divided across national and religious boundaries. Similarly, partisan caricature served as a rhetorical device to anticipate and challenge ideologically rigid interpretations. Together, such practices help construct the cross-border relationships through which a fragmented public can momentarily recognize itself as connected. However, when creators avoid political or religious topics, alter their presentation to appear sufficiently culturally authentic, or anticipate accusations of national disloyalty, it reproduces the boundaries that relational labor seeks to bridge. Importantly, as creators in postcolonial contexts learn which combinations of speaker, topic, language, and political position are likely to be accepted, and as contentious perspectives become less frequently produced, audiences encounter more narratives that already conform to dominant expectations, thereby shaping a recursive process that defeats its decolonial objective. We do not imply that YouTubers are responsible for that futility. Their individual strategies for protecting essential income, reputation, physical safety, or political security can narrow the collective outcomes.

This tension brings postcolonial and decolonial perspectives into closer dialogue~\cite{bhambra2014postcolonial, das2022collaborative}. A postcolonial analysis explains how inherited classifications of religion, language, gender, class, and nationality structure how speakers are recognized. In contrast, a decolonial analysis asks whether creators can contest those classifications and reclaim epistemic authority. Our paper shows that these processes cannot be separated in practice. Attempts to reclaim narrative authority take place through negotiations with publics that continue to enforce historically produced boundaries. This extends relational labor beyond relationship maintenance within affective publics and boundary policing~\cite{papacharissi2015affective, marwick2013status} by showing how such processes are intensified in postcolonial settings. This complicates dominant narratives of participatory culture that emphasize empowerment and connection~\cite{baym2012socially}, conceptualizing the relational work as the continuous work of becoming recognizable as a legitimate speaker while challenging the sociohistorically shaped terms of that recognition.

Across the contexts in which identities and publics have been colonially fragmented, digital participation can require people to become recognizable within historically produced categories while simultaneously contesting those categories and reclaiming narrative authority~\cite{das2022collaborative, das2024reimagining}. However, our findings in the Bengali context are not directly generalizable across postcolonial contexts, but provide a historicist attentive lens~\cite{soden2021time} for examining how the conditions of discursive sustainability vary across them~\cite{das2022collaborative}. For example, while increased participation and visibility among the Kurdish diaspora through social media have transformed Kurdish national identity into a more participatory but internally pluralistic imagined community, they have not produced a unified national public~\cite{aghapouri2020towards, fischer202114}. Despite the resemblance in how Bengali and Kurdish identities are divided by different political histories and understandings of belonging, the geopolitical stakes in these two contexts differ significantly. YouTubers in Bengali decolonial discourse negotiate legitimacy within established nation-states that recognize their ethnolinguistic identity (e.g., Bangladesh means the land of Bengalis, and Bengali is the official language there and in two Indian states), whereas Kurdish digital discourse articulates collective nationhood among a stateless people. Where Bengali creators may moderate contentious expression to remain legible within competing national publics, asserting Kurdish collective identity can itself constitute contentious political expression. The comparison suggests that historically fragmented publics may require similar relational and boundary work, while the form and consequences of that work depend on whether creators are negotiating belonging across existing structures or imagining something without one.

\subsection{Global Platform Policies' Postcolonial Misalignment}
Platform governance is usually examined in terms of moderation, monetization, and algorithmic visibility~\cite{gillespie2010politics, srnicek2017platform, burgess2018youtube}. The problem is not only that globally designed policies fail to reflect local circumstances~\cite{irani2010postcolonial}, but platform governance converts existing geopolitical and institutional inequalities into signals that directly organize cultural production.

A view from the Global North is financially worth more than a view from the Global South, or an MCN-affiliated channel may possess greater authority in a copyright dispute than the creator who originally produced the material--such arrangements stratify which audiences are economically valuable and whose cultural claims are practically defensible. Monetization~\cite{duffy2017not, joseph2024advertising}, copyright~\cite{fiesler2015understanding, fiesler2020lawful}, and moderation~\cite{jiang2023trade} on online platforms are often structured around economic value, formalized legal frameworks, and opaque enforcement processes.

Differential monetization can reorganize representation without explicitly moderating content. Creators are incentivized to address comparatively privileged diasporic viewers even when their cultural commitments or larger audiences remain in South Asia, illustrating the former's ``diasporic superposition"~\cite{das2026global}. This can give disproportionate visibility to selective diasporic understandings of Bengali culture while making locally changing, rural, or agrarian cultural practices less economically sustainable. Thus, platform capitalism influences which versions of a transnational culture can be produced often enough to become prominent. YouTubers accepted copyright and revenue losses to make audience-requested reaction videos, treated engagement and affinity as longer-term forms of value, sought direct support, and distributed content across several platforms.

Copyright enforcement is also experienced as inconsistent and mediated by institutional asymmetries, such as the influence of MCNs, which can amplify some creators' claims while undermining others~\cite{lobato2016cultural, siciliano2023intermediaries}. At the same time, moderation opacity introduces uncertainty about content visibility and accessibility, making outcomes difficult to anticipate and hence to agree on~\cite{das2021jol}. While collective folk theories help creators act under opacity~\cite{eslami2016first, karizat2021algorithmic}, this shifts responsibility for understanding enforcement from the governing institution to those subject to it. As a result, governance is not experienced as a stable or universally legible system~\cite{shahid2023decolonizing}, but as a set of unevenly enforced mechanisms that require continuous interpretation.

Overall, platform governance's postcolonial misalignment is not simply a failure to localize otherwise universal policies. It emerges through how users in postcolonial contexts adapt, reinterpret, and sometimes resist imposed technological structures. Therefore, sustaining decolonial discourse requires not only engaging with platform rules but also continuously reconciling the gaps between those rules and the lived realities of creators and their audiences.

\subsection{Risk, Trust, and Safety in Postcolonial Participation}
Participation in decolonial discourse is shaped by ongoing negotiations of risk and safety. Beyond harassment and safety concerns in online spaces documented by prior scholarship~\cite{marwick2017media, jhaver2019human}, these dynamics are intensified and reconfigured in postcolonial contexts, where identity, politics, and legal structures intersect. For our participants, visibility was not only an opportunity for engagement but also a source of vulnerability. They faced risks of broader sociopolitical and legal consequences for engaging with sensitive topics. These risks were often anticipatory rather than reactive. Even in the absence of direct intervention, the possibility of state attention, legal action, or institutional scrutiny shaped how creators calibrated their participation. Thus, risk operates not only through explicit enforcement but also through the anticipation of potential consequences, conditioning what can be expressed, how it is framed, and whether it is expressed at all.

These dynamics extend beyond individual creators to their collaborators, social networks, and production processes. The perceived risk of association with politically sensitive content creates barriers to collaboration, limits access to resources, and constrains forms of storytelling and reporting. Such conditions reflect what has been described as chilling effects~\cite{schauer1978fear}, where the threat of legal or political repercussions discourages participation even without direct enforcement. As a result, content creators may resort to indirect expression, avoid certain topics altogether, or limit the scope of their engagement. Trust plays a central role in mediating participation under these conditions. YouTubers described actively working to build and maintain trust with their audiences through linguistic alignment, cultural signaling, and consistent engagement, while also navigating distrust rooted in fragmented postcolonial identities~\cite{chatterjee1993nation}.

Thus, trust is not given but continuously negotiated, and remains fragile in environments where audiences may question motives, authenticity, or allegiance. This aligns with and extends prior work on trust and credibility in online communities~\cite{metzger2010social, niu2023building} by showing how trust is entangled with postcolonial identity politics, transnational audience formations, and conditions of uncertainty. Safety, in turn, is unevenly distributed and structurally constrained. Female creators and those engaging with contentious sociopolitical issues faced heightened exposure to both social and political risks, shaping their mobility, content choices, and modes of engagement. This aligns with prior social computing scholarship~\cite{rifat2024politics} showing how postcolonial memory shapes a politics of fear that constrains online participation among users from religious minority communities. Moreover, the absence of clear protections or recourse mechanisms within platform and institutional structures further exacerbates these vulnerabilities. Overall, decolonial discourse is shaped not only by what creators wish to express but also by what can be safely articulated within environments marked by anticipatory fear, legal uncertainty, and unequal exposure to harm.
\section{Limitations, Future Work, and Ethical Considerations}
While this paper examines how decolonial discourse is sustained under layered sociotechnical constraints in the context of Bengali sociopolitics and culture, we did not look at the algorithmic side of the platforms where these discourses take place. Sociotechnical systems (e.g., recommendation algorithms) can present unique politics by perpetuating algorithmic coloniality or hyper-nationalism~\cite{das2021jol}. In future work, we plan to examine how algorithmic systems further shape the visibility, reach, and viability of decolonial discourse, as well as how creators navigate these dynamics. Another limitation of our study is the lack of gender diversity among participants. Similar to prior work in Bangladeshi contexts~\cite{das2022understanding, das2024reimagining}, we found fewer female YouTubers producing videos within the sociopolitical decolonial discourse space, possibly due to the region's tense political environment~\cite{jalal2009democracy}. We recruited a few creators from religious minority backgrounds. Beyond the niche nature of local culture content, prior work shows that Bangladeshi minorities' social media participation is shaped by fear and a spiral of silence~\cite{rifat2024politics}. Despite multiple attempts, we were unable to recruit a Hindu participant from Bangladesh and included only one Muslim participant from India. The first author, as a religious minority, remained attentive to these dynamics, given that relations between majority and minority religions intersect with political power in South Asia.

Moreover, due to the conservative subcontinental culture, despite several focused attempts to recruit more female participants, we were only able to interview four female YouTubers. Among them, two participants requested to be interviewed alongside their YouTube channel's co-patrons on the same call (one cited following Muslim guidelines for women socializing with non-familial men as the rationale for her request). While interviews in a group setting created the possibility that some participants might suppress their opinions or that one participant might dominate the conversation, we did not observe any visible hesitancy. The interviewer also strategically navigated the conversation so that all participants in those calls were equally responsive.

Again, all but one of our participants had at least a university degree, which raises the possibility that the study may reflect the views of more highly educated people rather than the general population. While all our participants belong to previously colonized communities, they are from age groups that have not experienced British or Pakistani colonial rule themselves. Therefore, future work should look into understanding the experiences of people who not only belong to previously colonized communities but also experienced colonial subjugation themselves or faced colonially created crises (e.g., being refugees due to partition or war). When colonization is viewed as a crisis~\cite{das2022collaborative}, researchers should also consider the risk that older participants will relive traumatic experiences of the colonial past. Since our participants drew on experiences of living in colonially marginalized communities and the ones they heard from older family members, their risk of reliving such traumatic experiences was minimal. Moreover, our participant recruitment was heavily influenced by our search for relevant YouTube videos and channels.
\section{Conclusion}
This paper examined how YouTubers sustain Bengali decolonial discourse within a layered sociotechnical environment shaped by audience dynamics, platform governance, and state-linked regulation. Showing that such discourse is not merely content creation but a situated, collective practice, our findings underscores the need to move beyond platform-centric accounts of participation toward an integrated perspective that accounts for how historically rooted tensions, transnational audience formations, and uneven governance structures shape the viability of online expression. This highlights that supporting discourse in an ethnolinguistic group living within postcolonial contexts requires attention not only to interfaces and policies but also to the relational, political, and infrastructural conditions that make participation possible or fragile. More broadly, our work underscores that designing for global platforms entails engaging with the diverse and historically situated realities of their users, pointing toward more contextually grounded and equitable approaches to platform design and governance.

\bibliographystyle{ACM-Reference-Format}
\bibliography{acmart}

@article{gecas1982self,
  title={The self-concept},
  author={Gecas, Viktor},
  journal={Annual review of sociology},
  volume={8},
  number={1},
  pages={1--33},
  year={1982},
  publisher={Annual Reviews 4139 El Camino Way, PO Box 10139, Palo Alto, CA 94303-0139, USA}
}

@article{snow2001collective,
  title={Collective identity and expressive forms},
  author={Snow, David},
  year={2001},
  journal={International Encyclopedia of the Social \& Behavioral Sciences},
  pages={2212--2219},
  publisher={Center for the Study of Democracy}
}

@inproceedings{haimson2015disclosure,
  title={Disclosure, stress, and support during gender transition on Facebook},
  author={Haimson, Oliver L and Brubaker, Jed R and Dombrowski, Lynn and Hayes, Gillian R},
  booktitle={Proceedings of the 18th ACM Conference on Computer Supported Cooperative Work \& Social Computing},
  pages={1176--1190},
  year={2015}
}

@article{klassen2021more,
  title={More than a modern day Green book: Exploring the online community of Black Twitter},
  author={Klassen, Shamika and Kingsley, Sara and McCall, Kalyn and Weinberg, Joy and Fiesler, Casey},
  journal={Proc. of the ACM on Human-Computer Interaction},
  volume={5},
  number={},
  pages={},
  year={2021},
  publisher={ACM New York, NY, USA}
}

@article{soden2021time,
  title={Time for historicism in CSCW: An invitation},
  author={Soden, Robert and Ribes, David and Avle, Seyram and Sutherland, Will},
  journal={Proceedings of the ACM on Human-Computer Interaction},
  volume={5},
  number={CSCW2},
  pages={1--18},
  year={2021},
  publisher={ACM New York, NY, USA}
}

@inproceedings{irani2010postcolonial,
  title={Postcolonial computing: a lens on design and development},
  author={Irani, Lilly and Vertesi, Janet and Dourish, Paul and Philip, Kavita and Grinter, Rebecca E},
  booktitle={Proceedings of the SIGCHI conference on human factors in computing systems},
  pages={1311--1320},
  year={2010}
}

@book{fanon2008black,
  title={Black skin, white masks},
  author={Fanon, Frantz},
  year={2008},
  publisher={Grove press}
}

@book{lugones2016coloniality,
  title={The coloniality of gender},
  author={Lugones, Maria},
  year={2016},
  publisher={Springer}
}

@book{chatterjee1993nation,
  title={The nation and its fragments: Colonial and postcolonial histories},
  author={Chatterjee, Partha},
  year={1993},
  publisher={Princeton University Press}
}

@article{das2021jol,
  title={``Jol" or ``Pani"?: How Does Governance Shape a Platform's Identity?},
  author={Das, Dipto and {\O}sterlund, Carsten and Semaan, Bryan},
  journal={Proceedings of the ACM on Human-Computer Interaction},
  volume={5},
  number={CSCW2},
  pages={1--25},
  year={2021},
  publisher={ACM New York, NY, USA}
}

@article{das2024reimagining,
  title={Reimagining Communities through Transnational Bengali Decolonial Discourse with YouTube Content Creators},
  author={Das, Dipto and Gandhi, Dhwani and Semaan, Bryan},
  pages={1--35},
  year={2024},
  note={Under review at ACM Conference on Computer-Supported Cooperative Work and Social Computing}
}

@inproceedings{das2022collaborative,
  title={Collaborative identity decolonization as reclaiming narrative agency: Identity work of Bengali communities on Quora},
  author={Das, Dipto and Semaan, Bryan},
  booktitle={Proceedings of the 2022 CHI Conference on Human Factors in Computing Systems},
  pages={1--23},
  year={2022}
}

@article{laenui2000processes,
  title={Processes of decolonization},
  author={Laenui, Poka},
  journal={Reclaiming Indigenous voice and vision},
  pages={150--160},
  year={2000},
  publisher={Vancouver, BC: University of British Columbia Press}
}

@article{dosono2020decolonizing,
  title={Decolonizing tactics as collective resilience: Identity work of AAPI communities on {R}eddit},
  author={Dosono, Bryan and Semaan, Bryan},
  journal={Proceedings of the ACM on Human-Computer interaction},
  volume={4},
  number={CSCW1},
  pages={1--20},
  year={2020},
  publisher={ACM New York, NY, USA}
}

@article{jiang2023trade,
  title={A trade-off-centered framework of content moderation},
  author={Jiang, Jialun Aaron and Nie, Peipei and Brubaker, Jed R and Fiesler, Casey},
  journal={ACM Transactions on Computer-Human Interaction},
  volume={30},
  number={1},
  pages={1--34},
  year={2023},
  publisher={ACM New York, NY}
}

@book{fanon2007wretched,
  title={The Wretched of the Earth},
  author={Fanon, Frantz},
  year={2007},
  publisher={Grove Atlantic}
}

@inproceedings{das2023studying,
  title={Studying Multi-dimensional Marginalization of Identity from Decolonial and Postcolonial Perspectives},
  author={Das, Dipto},
  booktitle={Companion Publication of the 2023 Conference on Computer Supported Cooperative Work and Social Computing},
  pages={437--440},
  year={2023}
}

@inproceedings{schlesinger2017intersectional,
  title={Intersectional HCI: Engaging identity through gender, race, and class},
  author={Schlesinger, Ari and Edwards, W Keith and Grinter, Rebecca E},
  booktitle={Proceedings of the 2017 CHI conference on human factors in computing systems},
  pages={5412--5427},
  year={2017}
}

@article{liang2021embracing,
  title={Embracing four tensions in human-computer interaction research with marginalized people},
  author={Liang, Calvin A and Munson, Sean A and Kientz, Julie A},
  journal={ACM Transactions on Computer-Human Interaction (TOCHI)},
  volume={28},
  number={2},
  pages={1--47},
  year={2021},
  publisher={ACM New York, NY, USA}
}

@book{sen2018decline,
  title={The decline of the caste question: Jogendranath Mandal and the defeat of Dalit politics in Bengal},
  author={Sen, Dwaipayan},
  year={2018},
  publisher={Cambridge University Press}
}

@book{van2020history,
  title={A history of {B}angladesh},
  author={Van Schendel, Willem},
  year={2020},
  publisher={Cambridge University Press}
}

@book{pandey2001remembering,
  title={Remembering partition: Violence, nationalism and history in {I}ndia},
  author={Pandey, Gyanendra and others},
  volume={7},
  year={2001},
  publisher={Cambridge University Press}
}

@article{tajfel1974social,
  title={Social identity and intergroup behaviour},
  author={Tajfel, Henri},
  journal={Social science information},
  volume={13},
  number={2},
  pages={65--93},
  year={1974},
  publisher={Sage Publications Sage CA: Thousand Oaks, CA}
}

@book{anderson2006imagined,
  title={Imagined communities: Reflections on the origin and spread of nationalism},
  author={Anderson, Benedict},
  year={2006},
  publisher={Verso books}
}

@article{ma2021advertiser,
  title={" How advertiser-friendly is my video?": YouTuber's Socioeconomic Interactions with Algorithmic Content Moderation},
  author={Ma, Renkai and Kou, Yubo},
  journal={Proceedings of the ACM on Human-Computer Interaction},
  volume={5},
  number={CSCW2},
  year={2021},
  publisher={ACM New York, NY, USA}
}

@inproceedings{houston2016values,
  title={Values in repair},
  author={Houston, Lara and Jackson, Steven J and Rosner, Daniela K and Ahmed, Syed Ishtiaque and Young, Meg and Kang, Laewoo},
  booktitle={Proceedings of the 2016 CHI conference on human factors in computing systems},
  pages={1403--1414},
  year={2016}
}

@article{dym2019coming,
  title={" Coming Out Okay" Community Narratives for LGBTQ Identity Recovery Work},
  author={Dym, Brianna and Brubaker, Jed R and Fiesler, Casey and Semaan, Bryan},
  journal={Proceedings of the ACM on Human-Computer Interaction},
  volume={3},
  number={CSCW},
  pages={1--28},
  year={2019},
  publisher={ACM New York, NY, USA}
}

@article{ghoshal2020toward,
  title={Toward a grassroots culture of technology practice},
  author={Ghoshal, Sucheta and Mendhekar, Rishma and Bruckman, Amy},
  journal={Proceedings of the ACM on Human-Computer Interaction},
  volume={4},
  number={CSCW1},
  pages={1--28},
  year={2020},
  publisher={ACM New York, NY, USA}
}

@article{ibarra2010identity,
  title={Identity work and play},
  author={Ibarra, Herminia and Petriglieri, Jennifer L},
  journal={Journal of Organizational Change Management},
  volume={23},
  number={1},
  pages={10--25},
  year={2010},
  publisher={Emerald Group Publishing Limited}
}

@article{lange2007publicly,
  title={Publicly private and privately public: Social networking on YouTube},
  author={Lange, Patricia G},
  journal={Journal of computer-mediated communication},
  volume={13},
  number={1},
  pages={361--380},
  year={2007},
  publisher={Oxford University Press Oxford, UK}
}

@article{spivak2003can,
  title={Can the subaltern speak?},
  author={Spivak, Gayatri Chakravorty},
  journal={Die Philosophin},
  volume={14},
  number={27},
  pages={42--58},
  year={2003}
}

@misc{editors2022hindustani,
  title={{H}industani language},
  author={Encyclopaedia Britannica},
  year={2022},
  howpublished={\url{https://www.britannica.com/topic/Hindustani-language}},
  note={Online, Last accessed: December 21, 2022}
}

@article{rifat2024politics,
  title={The Politics of Fear and the Experience of {B}angladeshi Religious Minority Communities Using Social Media Platforms},
  author={Rifat, Mohammad Rashidujjaman and Das, Dipto and Podder, Arpon and Jannat, Mahiratul and Soden, Robert and Semaan, Bryan and Ahmed, Syed Ishtiaque},
  journal={Proceedings of the ACM on human-computer interaction},
  number={CSCW},
  pages={1--31},
  year={2024},
  publisher={ACM New York, NY, USA},
  note={in press}
}

@inproceedings{milliken2008user-sphere,
  title={User-generated online video: The next public sphere?},
  author={Milliken, Mary C and O'Donnell, Susan},
  booktitle={2008 IEEE International Symposium on Technology and Society},
  pages={1--3},
  year={2008},
  organization={IEEE}
}

@article{munoz2022platform,
  title={Platform-mediated Markets, Online Freelance Workers and Deconstructed Identities},
  author={Munoz, Isabel and Dunn, Michael and Sawyer, Steve and Michaels, Emily},
  journal={Proceedings of the ACM on Human-Computer Interaction},
  volume={6},
  number={CSCW2},
  year={2022},
  publisher={ACM New York, NY, USA}
}

@article{hall1989cultural,
  title={Cultural identity and cinematic representation},
  author={Hall, Stuart},
  journal={Journal of Cinema and Media},
  year={1989},
  publisher={JSTOR}
}

@article{askanius2014video,
  title={Video for change},
  author={Askanius, Tina},
  journal={The handbook of development communication and social change},
  year={2014},
  publisher={Wiley Online Library},
}

@article{goodman1961snowball,
  title={Snowball sampling},
  author={Goodman, Leo A},
  journal={The annals of mathematical statistics},
  pages={148--170},
  year={1961},
  publisher={JSTOR}
}

@book{nandy1989intimate,
  title={The Intimate Enemy: Loss and Recovery of Self Under Colonialism},
  author={Nandy, Ashis},
  year={1989},
  publisher={Oxford University Press Oxford}
}

@article{madan1972two,
  title={Two Faces of Bengali Ethnicity: Muslim Bengali or Bengali Muslim},
  author={Madan, Triloki Nath},
  journal={The developing economies},
  volume={10},
  number={1},
  pages={74--85},
  year={1972},
  publisher={Wiley Online Library}
}

@article{vaghela2021birds,
  title={Birds of a caste-how caste hierarchies manifest in retweet behavior of indian politicians},
  author={Vaghela, Palashi and K Mothilal, Ramaravind and Pal, Joyojeet},
  journal={Proceedings of the ACM on Human-Computer Interaction},
  volume={4},
  number={CSCW3},
  year={2021},
  publisher={ACM New York, NY, USA}
}

@book{chatterji2002bengal,
  title={Bengal divided: {H}indu communalism and partition, 1932-1947},
  author={Chatterji, Joya},
  year={2002},
  publisher={Cambridge University Press}
}

@inproceedings{morioka2016identity,
  title={Identity work on social media sites: Disadvantaged students' college transition processes},
  author={Morioka, Tsubasa and Ellison, Nicole B and Brown, Michael},
  booktitle={Proceedings of the 19th ACM conference on computer-supported cooperative work \& social computing},
  pages={848--859},
  year={2016}
}

@article{askanius2011online,
  title={Online social media for radical politics: climate change activism on YouTube},
  author={Askanius, Tina and Uldam, Julie},
  journal={International journal of electronic governance},
  volume={4},
  number={1-2},
  pages={69--84},
  year={2011},
  publisher={Inderscience Publishers}
}

@inproceedings{dash2022insights,
  title={Insights Into Incitement: A Computational Perspective on Dangerous Speech on Twitter in {I}ndia},
  author={Dash, Saloni and Grover, Rynaa and Shekhawat, Gazal and Kaur, Sukhnidh and Mishra, Dibyendu and Pal, Joyojeet},
  booktitle={ACM SIGCAS/SIGCHI Conference on Computing and Sustainable Societies (COMPASS)},
  pages={103--121},
  year={2022}
}

@article{gillespie2010politics,
  title={The politics of ‘platforms’},
  author={Gillespie, Tarleton},
  journal={New media \& society},
  volume={12},
  number={3},
  pages={347--364},
  year={2010},
  publisher={SAGE Publications Sage UK: London, England}
}

@article{barta2021constructing,
  title={Constructing Authenticity on TikTok: Social Norms and Social Support on the" Fun" Platform},
  author={Barta, Kristen and Andalibi, Nazanin},
  journal={Proceedings of the ACM on Human-Computer Interaction},
  volume={5},
  number={CSCW2},
  pages={1--29},
  year={2021},
  publisher={ACM New York, NY, USA}
}

@book{jalal2009democracy,
  title={Democracy and authoritarianism in South Asia},
  author={Jalal, Ayesha},
  year={2009},
  publisher={Cambridge University Press}
}

@book{burgess2018youtube,
  title={YouTube: Online video and participatory culture},
  author={Burgess, Jean and Green, Joshua},
  year={2018},
  publisher={John Wiley \& Sons}
}

@inproceedings{das2022note,
  title={Note: A Sociomaterial Perspective on Trace Data Collection: Strategies for Democratizing and Limiting Bias},
  author={Das, Dipto and Podder, Arpon and Semaan, Bryan},
  booktitle={ACM SIGCAS/SIGCHI Conference on Computing and Sustainable Societies (COMPASS)},
  pages={569--573},
  year={2022}
}

@inproceedings{das2022understanding,
  title={Understanding the Strategies and Practices of Facebook Microcelebrities for Engaging in Sociopolitical Discourses},
  author={Das, Dipto and Islam, A.K.M. Najmul and Haque, S.M. Taiabul and Vuorinen, Jukka and Ahmed, Syed Ishtiaque},
  booktitle={International Conference on Information \& Communication Technologies and Development},
  pages={1--19},
  year={2022}
}

@article{strauss1994grounded,
  title={Grounded theory methodology},
  author={Strauss, Anselm and Corbin, Juliet},
  journal={Handbook of qualitative research},
  volume={17},
  number={1},
  pages={273--285},
  year={1994},
  publisher={Thousand Oaks, CA}
}

@article{chen2023my,
  title={" My Culture, My People, My Hometown": Chinese Ethnic Minorities Seeking Cultural Sustainability by Video Blogging},
  author={Chen, Si and Chen, Xinyue and Lu, Zhicong and Huang, Yun},
  journal={Proceedings of the ACM on Human-Computer Interaction},
  volume={7},
  number={CSCW1},
  pages={1--30},
  year={2023},
  publisher={ACM New York, NY, USA}
}

@book{erikson1968identity,
  title={Identity: Youth and crisis},
  author={Erikson, Erik H},
  volume={7},
  year={1968},
  publisher={WW Norton \& company}
}

@article{vaghela2022caste,
  title={Caste Capital on Twitter: A Formal Network Analysis of Caste Relations among {I}ndian Politicians},
  author={Vaghela, Palashi and Mothilal, Ramaravind Kommiya and Romero, Daniel and Pal, Joyojeet},
  journal={Proc. of the ACM on HCI},
  year={2022},
  publisher={ACM New York, NY, USA}
}

@incollection{sinha2017colonial,
  title={Colonial masculinity: The ‘manly Englishman’and the ‘effeminate Bengali’in the late nineteenth century},
  author={Sinha, Mrinalini},
  booktitle={Colonial masculinity},
  year={2017},
  publisher={Manchester University Press}
}

@book{goffman1978presentation,
  title={The presentation of self in everyday life},
  author={Goffman, Erving},
  year={1978},
  publisher={Harmondsworth London}
}

@inproceedings{dosono2018identity,
  title={Identity work as deliberation: AAPI political discourse in the 2016 US Presidential Election},
  author={Dosono, Bryan and Semaan, Bryan},
  booktitle={Proceedings of the 2018 CHI Conference on Human Factors in Computing Systems},
  pages={1--12},
  year={2018}
}

@inproceedings{cha2007tube,
  title={I tube, you tube, everybody tubes: analyzing the world's largest user generated content video system},
  author={Cha, Meeyoung and Kwak, Haewoon and Rodriguez, Pablo and Ahn, Yong-Yeol and Moon, Sue},
  booktitle={Proceedings of the 7th ACM SIGCOMM conference on Internet measurement},
  pages={1--14},
  year={2007}
}

@article{kommiya2022voting,
  title={Voting with the Stars: Analyzing Partisan Engagement between Celebrities and Politicians in {I}ndia},
  author={Kommiya Mothilal, Ramaravind and Mishra, Dibyendu and Nishal, Sachita and Lalani, Faisal M and Pal, Joyojeet},
  journal={Proceedings of the ACM on Human-Computer Interaction},
  volume={6},
  number={CSCW1},
  pages={1--29},
  year={2022},
  publisher={ACM New York, NY, USA}
}

@misc{yuan2022what,
  title={What is Creator Economy?},
  author={Yuan, Yuanling and Constine, Josh},
  year={2022},
  howpublished={\url{https://signalfire.com/blog/creator-economy/}},
  note={last accessed: September 20, 2022}
}

@article{bhambra2014postcolonial,
  title={Postcolonial and decolonial dialogues},
  author={Bhambra, Gurminder K},
  journal={Postcolonial studies},
  volume={17},
  number={2},
  pages={115--121},
  year={2014},
  publisher={Taylor \& Francis}
}

@article{jhaver2019human,
  title={Human-machine collaboration for content regulation: The case of reddit automoderator},
  author={Jhaver, Shagun and Birman, Iris and Gilbert, Eric and Bruckman, Amy},
  journal={ACM Transactions on Computer-Human Interaction (TOCHI)},
  volume={26},
  number={5},
  pages={1--35},
  year={2019},
  publisher={ACM New York, NY, USA}
}

@misc{yin2017case,
  title={Case study research and applications: Design and methods},
  author={Yin Robert, K},
  year={2017},
  publisher={Sage publications Thousand Oaks, CA}
}

@article{anderson1994representations,
  title={Representations and requirements: The value of ethnography in system design},
  author={Anderson, Robert J},
  journal={Human-computer interaction},
  volume={9},
  number={2},
  pages={151--182},
  year={1994},
  publisher={Taylor \& Francis}
}

@misc{wiki2022hindi,
  title={Hindi–Urdu controversy},
  author={Wikipedia contributors},
  year={2022},
  howpublished={\url{https://en.wikipedia.org/wiki/Hindi-Urdu_controversy}},
  note={Online; Last accessed December 21, 2022}
}

@article{suri2011purposeful,
  title={Purposeful sampling in qualitative research synthesis},
  author={Suri, Harsh},
  journal={Qualitative research journal},
  volume={11},
  number={2},
  pages={63--75},
  year={2011},
  publisher={Emerald Group Publishing Limited}
}

@article{mcdonald2019reliability,
  title={Reliability and inter-rater reliability in qualitative research: Norms and guidelines for CSCW and HCI practice},
  author={McDonald, Nora and Schoenebeck, Sarita and Forte, Andrea},
  journal={Proceedings of the ACM on Human-Computer Interaction},
  volume={3},
  number={CSCW},
  pages={1--23},
  year={2019},
  publisher={ACM New York, NY, USA}
}

@inproceedings{al2010blogging,
  title={Blogging in a region of conflict: supporting transition to recovery},
  author={Al-Ani, Ban and Mark, Gloria and Semaan, Bryan},
  booktitle={Proceedings of the SIGCHI Conference on human factors in computing systems},
  pages={1069--1078},
  year={2010}
}

@inproceedings{shahid2023decolonizing,
  title={Decolonizing Content Moderation: Does Uniform Global Community Standard Resemble Utopian Equality or Western Power Hegemony?},
  author={Shahid, Farhana and Vashistha, Aditya},
  booktitle={Proceedings of the 2023 CHI Conference on Human Factors in Computing Systems},
  pages={1--18},
  year={2023}
}

@article{leung2015validity,
  title={Validity, reliability, and generalizability in qualitative research},
  author={Leung, Lawrence},
  journal={Journal of family medicine and primary care},
  volume={4},
  number={3},
  pages={324},
  year={2015},
  publisher={Wolters Kluwer--Medknow Publications}
}

@article{simpson2021you,
  title={For You, or For" You"? Everyday LGBTQ+ Encounters with TikTok},
  author={Simpson, Ellen and Semaan, Bryan},
  journal={Proceedings of the ACM on human-computer interaction},
  volume={4},
  number={CSCW3},
  pages={1--34},
  year={2021},
  publisher={ACM New York, NY, USA}
}

@book{wengraf2001qualitative,
  title={Qualitative research interviewing: Biographic narrative and semi-structured methods},
  author={Wengraf, Tom},
  year={2001},
  publisher={sage}
}

@inproceedings{seth2022closed,
  title={Closed Ranks: The Discursive Value of Military Support for {I}ndian Politicians on Social Media},
  author={Seth, Agrima and De, Soham and Arya, Arshia and Wilkinson, Steven and Singh, Sushant and Pal, Joyojeet},
  booktitle={Proceedings of the 2022 International Conference on Information and Communication Technologies and Development},
  pages={1--11},
  year={2022}
}

@inproceedings{irani2009postcolonial,
  title={Postcolonial interculturality},
  author={Irani, Lilly C and Dourish, Paul},
  booktitle={Proceedings of the 2009 international workshop on Intercultural collaboration},
  pages={249--252},
  year={2009}
}

@misc{globalmediainsight2023YouTubeStatistics,
	author = {Global Media Insight},
	title = {{Y}ou{T}ube {S}tatistics 2023 [{U}sers by {C}ountry + {D}emographics] --- globalmediainsight.com},
	howpublished = {\url{https://www.globalmediainsight.com/blog/youtube-users-statistics/}},
	year = {2023},
	note = {[Accessed 12-Jul-2023]},
}

@article{fiesler2019creativity,
  title={Creativity, copyright, and close-knit communities: A case study of social norm formation and enforcement},
  author={Fiesler, Casey and Bruckman, Amy S},
  journal={Proceedings of the ACM on Human-Computer Interaction},
  volume={3},
  number={GROUP},
  pages={1--24},
  year={2019},
  publisher={ACM New York, NY, USA}
}

@inproceedings{sharma2017analyzing,
  title={Analyzing ideological discourse on social media: A case study of the abortion debate},
  author={Sharma, Eva and Saha, Koustuv and Ernala, Sindhu Kiranmai and Ghoshal, Sucheta and De Choudhury, Munmun},
  booktitle={Proceedings of the 2017 international conference of the computational social science society of the americas},
  pages={1--8},
  year={2017}
}

@article{kumar2018uber,
  title={Uber in {B}angladesh: The Tangled Web of mobility and justice},
  author={Kumar, Neha and Jafarinaimi, Nassim and Bin Morshed, Mehrab},
  journal={Proceedings of the ACM on Human-Computer Interaction},
  volume={2},
  number={CSCW},
  pages={1--21},
  year={2018},
  publisher={ACM New York, NY, USA}
}

@article{karizat2021algorithmic,
  title={Algorithmic folk theories and identity: How TikTok users co-produce Knowledge of identity and engage in algorithmic resistance},
  author={Karizat, Nadia and Delmonaco, Dan and Eslami, Motahhare and Andalibi, Nazanin},
  journal={Proceedings of the ACM on Human-Computer Interaction},
  volume={5},
  number={CSCW2},
  pages={1--44},
  year={2021},
  publisher={ACM New York, NY, USA}
}

@article{fiesler2023chilling,
  title={Chilling Tales: Understanding the Impact of Copyright Takedowns on Transformative Content Creators},
  author={Fiesler, Casey and Paup, Joshua and Zacher, Corian},
  journal={Proceedings of the ACM on Human-Computer Interaction},
  volume={7},
  number={CSCW2},
  pages={1--21},
  year={2023},
  publisher={ACM New York, NY, USA}
}

@inproceedings{lu2019feel,
  title={" I feel it is my responsibility to stream" Streaming and Engaging with Intangible Cultural Heritage through Livestreaming},
  author={Lu, Zhicong and Annett, Michelle and Fan, Mingming and Wigdor, Daniel},
  booktitle={Proceedings of the 2019 CHI Conference on Human Factors in Computing Systems},
  pages={1--14},
  year={2019}
}

@article{milliken2008user-canada,
  title={User-generated video and the online public sphere: Will YouTube facilitate digital freedom of expression in Atlantic Canada},
  author={Milliken, Mary and Gibson, Kerri and O’Donnell, Susan},
  journal={American Communication Journal},
  volume={10},
  number={3},
  year={2008},
  publisher={Citeseer}
}

@inproceedings{panda2020affording,
  title={Affording extremes: incivility, social media and democracy in the {I}ndian context},
  author={Panda, Anmol and Chakraborty, Sunandan and Raval, Noopur and Zhang, Han and Mohapatra, Mugdha and Akbar, Syeda Zainab and Pal, Joyojeet},
  booktitle={Proceedings of the 2020 International Conference on Information and Communication Technologies and Development},
  pages={1--12},
  year={2020}
}

@article{sebastian2019distinguishing,
  title={Distinguishing between the strains grounded theory: Classical, interpretive and constructivist},
  author={Sebastian, Kailah},
  journal={Journal for Social Thought},
  volume={3},
  number={1},
  year={2019}
}

@inproceedings{mcroberts2016viewers,
  title={Do it for the viewers! Audience engagement behaviors of young YouTubers},
  author={McRoberts, Sarah and Bonsignore, Elizabeth and Peyton, Tamara and Yarosh, Svetlana},
  booktitle={Proc. of International Conference on Interaction Design and Children},
  year={2016}
}

@article{lobato2016cultural,
  title={The cultural logic of digital intermediaries: YouTube multichannel networks},
  author={Lobato, Ramon},
  journal={Convergence},
  volume={22},
  number={4},
  pages={348--360},
  year={2016},
  publisher={Sage Publications Sage UK: London, England}
}

@book{srnicek2017platform,
  title={Platform capitalism},
  author={Srnicek, Nick},
  year={2017},
  publisher={John Wiley \& Sons}
}

@incollection{zuboff2023age,
  title={The age of surveillance capitalism},
  author={Zuboff, Shoshana},
  booktitle={Social theory re-wired},
  pages={203--213},
  year={2023},
  publisher={Routledge}
}

@inproceedings{simpson2023rethinking,
  title={Rethinking creative labor: A sociotechnical examination of creativity \& creative work on TikTok},
  author={Simpson, Ellen and Semaan, Bryan},
  booktitle={Proceedings of the 2023 CHI Conference on Human Factors in Computing Systems},
  pages={1--16},
  year={2023}
}

@inproceedings{fiesler2015understanding,
  title={Understanding copyright law in online creative communities},
  author={Fiesler, Casey and Feuston, Jessica L and Bruckman, Amy S},
  booktitle={Proceedings of the 18th ACM conference on computer supported cooperative work \& social computing},
  pages={116--129},
  year={2015}
}

@incollection{fiesler2014copyright,
  title={Copyright terms in online creative communities},
  author={Fiesler, Casey and Bruckman, Amy},
  booktitle={CHI'14 Extended Abstracts on Human Factors in Computing Systems},
  pages={2551--2556},
  year={2014},
  publisher={ACM New York, NY, USA}
}

@inproceedings{fiesler2020lawful,
  title={Lawful users: Copyright circumvention and legal constraints on technology use},
  author={Fiesler, Casey},
  booktitle={Proceedings of the 2020 CHI Conference on Human Factors in Computing Systems},
  pages={1--11},
  year={2020}
}

@article{siciliano2023intermediaries,
  title={Intermediaries in the age of platformized gatekeeping: The case of YouTube “creators” and MCNs in the US},
  author={Siciliano, Michael L},
  journal={Poetics},
  volume={97},
  pages={101748},
  year={2023},
  publisher={Elsevier}
}

@article{graves2016law,
  title={The Law of YouTubers: The Next Generation of Creators and the Legal Issues They Face},
  author={Graves, Franklin and Lee, Michael},
  journal={Landslide},
  volume={9},
  pages={8},
  year={2016},
  publisher={HeinOnline}
}

@misc{uwnddmca,
  title={Digital Millennium Copyright Act (DMCA)},
  author={UW Copyright Resources},
  year={n.d.},
  howpublished={\url{https://copyrightresource.uw.edu/copyright-law/dmca/}},
  note={Last accessed: June 23, 2024} 
}

@article{perel2015accountability,
  title={Accountability in algorithmic copyright enforcement},
  author={Perel, Maayan and Elkin-Koren, Niva},
  journal={Stan. Tech. L. Rev.},
  volume={19},
  pages={473},
  year={2015},
  publisher={HeinOnline}
}

@article{das2026global,
  title={How do the Global South Diasporas Mobilize for Transnational Political Change?},
  author={Das, Dipto and Prio, Afrin and Saha, Pritu and Guha, Shion and Ahmed, Syed Ishtiaque},
  journal={arXiv preprint arXiv:2601.12705},
  year={2026}
}

@book{warner2021publics,
  title={Publics and counterpublics},
  author={Warner, Michael},
  year={2021},
  publisher={Princeton University Press}
}

@article{lindtner2011towards,
  title={Towards a framework of publics: Re-encountering media sharing and its user},
  author={Lindtner, Silvia and Chen, Judy and Hayes, Gillian R and Dourish, Paul},
  journal={ACM Transactions on Computer-Human Interaction (TOCHI)},
  volume={18},
  number={2},
  pages={1--23},
  year={2011},
  publisher={ACM New York, NY, USA}
}

@article{baym2012socially,
  title={Socially mediated publicness: An introduction},
  author={Baym, Nancy K and Boyd, Danah},
  journal={Journal of broadcasting \& electronic media},
  volume={56},
  number={3},
  pages={320--329},
  year={2012},
  publisher={Taylor \& Francis}
}

@article{marwick2011tweet,
  title={I tweet honestly, I tweet passionately: Twitter users, context collapse, and the imagined audience},
  author={Marwick, Alice E and Boyd, Danah},
  journal={New media \& society},
  volume={13},
  number={1},
  pages={114--133},
  year={2011},
  publisher={Sage Publications Sage UK: London, England}
}

@book{marwick2013status,
  title={Status update: Celebrity, publicity, and branding in the social media age},
  author={Marwick, Alice E},
  year={2013},
  publisher={yale university press}
}

@book{papacharissi2015affective,
  title={Affective publics: Sentiment, technology, and politics},
  author={Papacharissi, Zizi},
  year={2015},
  publisher={Oxford University Press}
}

@inproceedings{das2023decolonization,
  title={Decolonization through technology and decolonization of technology},
  author={Das, Dipto},
  booktitle={Companion Proceedings of the 2023 ACM International Conference on Supporting Group Work},
  pages={51--53},
  year={2023}
}

@phdthesis{das2024identity,
  title={Identity Decolonization amid the Coloniality of Computing},
  author={Das, Dipto},
  year={2024},
  school={University of Colorado at Boulder}
}

@article{klassen2022black,
  title={Black Lives, Green Books, and Blue Checks: Comparing the Content of the Negro Motorist Green Book to the Content on Black Twitter},
  author={Klassen, Shamika and Kingsley, Sara and McCall, Kalyn and Weinberg, Joy and Fiesler, Casey},
  journal={Proceedings of the ACM on Human-Computer Interaction},
  volume={6},
  number={GROUP},
  pages={1--22},
  year={2022},
  publisher={ACM New York, NY, USA}
}

@inproceedings{lingel2014city,
  title={City, self, network: transnational migrants and online identity work},
  author={Lingel, Jessica and Naaman, Mor and Boyd, Danah M},
  booktitle={Proceedings of the 17th ACM conference on Computer supported cooperative work \& social computing},
  pages={1502--1510},
  year={2014}
}

@inproceedings{dosono2017exploring,
  title={Exploring AAPI identity online: Political ideology as a factor affecting identity work on Reddit},
  author={Dosono, Bryan and Semaan, Bryan and Hemsley, Jeff},
  booktitle={Proceedings of the 2017 CHI conference extended abstracts on human factors in computing systems},
  pages={2528--2535},
  year={2017}
}

@incollection{mukhongodecolonizing,
  title={Decolonizing digital hegemonies: Reframing, disrupting, and occupying online spaces},
  author={Mukhongo, L Lusike},
  booktitle={Decolonising Approaches to Users and Audiences in the Global South},
  pages={88--104},
  publisher={Routledge}
}

@article{harris2023honestly,
  title={" Honestly, I Think TikTok has a Vendetta Against Black Creators": Understanding Black Content Creator Experiences on TikTok},
  author={Harris, Camille and Johnson, Amber Gayle and Palmer, Sadie and Yang, Diyi and Bruckman, Amy},
  journal={Proceedings of the ACM on Human-Computer Interaction},
  volume={7},
  number={CSCW2},
  pages={1--31},
  year={2023},
  publisher={ACM New York, NY, USA}
}

@article{kumar2019algorithmic,
  title={The algorithmic dance: YouTube's Adpocalypse and the gatekeeping of cultural content on digital platforms},
  author={Kumar, Sangeet},
  journal={Internet Policy Review},
  volume={8},
  number={2},
  pages={1--21},
  year={2019},
  publisher={Berlin: Alexander von Humboldt Institute for Internet and Society}
}

@article{joseph2024advertising,
  title={Advertising as governance: The digital commodity audience and platform advertising dependency},
  author={Joseph, Daniel and Bishop, Sophie},
  journal={Media, Culture \& Society},
  volume={46},
  number={6},
  pages={1269--1286},
  year={2024},
  publisher={SAGE Publications Sage UK: London, England}
}

@book{duffy2017not,
  title={(Not) getting paid to do what you love: Gender, social media, and aspirational work},
  author={Duffy, Brooke Erin},
  year={2017},
  publisher={Yale University Press}
}

@misc{ahmed2025youtube,
  title={YouTube CPM Rates in 2025: How Location Shapes Earnings},
  author={Ahmed, Irfan},
  howpublished={\url{https://www.digitalinformationworld.com/2025/08/youtube-cpm-rates-in-2025-how-location.html}},
  year={2025},
  note={[Accessed: March 27, 2026]}
}

@article{childs2023examining,
  title={Examining the production of co-active channels on youtube and bitchute},
  author={Childs, Matthew C and Horne, Benjamin D},
  journal={arXiv preprint arXiv:2303.07861},
  year={2023}
}

@article{van2023investigating,
  title={Investigating the Impacts of YouTube's Content Policies on Journalism and Political Discourse},
  author={Van Natta, Jared and Masadeh, Saleem and Hamilton, Bill},
  journal={Proceedings of the ACM on Human-Computer Interaction},
  volume={7},
  number={CSCW1},
  pages={1--28},
  year={2023},
  publisher={ACM New York, NY, USA}
}

@inproceedings{pisharody2025changes,
  title={Changes in YouTube's Content Moderation Policy Had Little Detectable Impact on Election Denial Content},
  author={Pisharody, Nilima and Norton, Sean T and Greene, Kevin T and Shapiro, Jacob N},
  booktitle={Proceedings of the International AAAI Conference on Web and Social Media},
  volume={19},
  pages={1550--1573},
  year={2025}
}

@article{hosseinmardi2021examining,
  title={Examining the consumption of radical content on YouTube},
  author={Hosseinmardi, Homa and Ghasemian, Amir and Clauset, Aaron and Mobius, Markus and Rothschild, David M and Watts, Duncan J},
  journal={Proceedings of the national academy of sciences},
  volume={118},
  number={32},
  pages={e2101967118},
  year={2021},
  publisher={National Academy of Sciences}
}

@article{vogele2017s,
  title={Where's the fair use: The takedown of let's play and reaction videos on YouTube and the need for comprehensive DMCA reform},
  author={Vogele, Jessica},
  journal={Touro L. Rev.},
  volume={33},
  pages={589},
  year={2017},
  publisher={HeinOnline}
}

@article{xiao2025institutionalizing,
  title={Institutionalizing Folk Theories of Algorithms: How Multi-Channel Networks (MCNs) Govern Algorithmic Labor in Chinese Live-Streaming Industry},
  author={Xiao, Qing and Chen, Rongyi and Xiao, Jingjia and Fu, Tianyang and Zhang, Alice Qian and Fan, Xianzhe and Zhang, Bingbing and Lu, Zhicong and Shen, Hong},
  journal={arXiv preprint arXiv:2505.20623},
  year={2025}
}

@misc{allndtier,
  title={Tier A MCN},
  author={All Content},
  howpublished={\url{https://www.allcontent.net/youtube-monetization-optimization-details/34/tier-a-mcn/en}},
  year={n.d.},
  note={[Accessed: March 27, 2026]}
}

@article{fasel2025between,
  title={Between regulation, pressure and collaboration: the public--private entanglement in content moderation},
  author={Fasel, Mathieu and Weerts, Sophie},
  journal={Telecommunications Policy},
  pages={103024},
  year={2025},
  publisher={Elsevier}
}

@book{gorwa2024politics,
  title={The politics of platform regulation: How governments shape online content moderation},
  author={Gorwa, Robert},
  year={2024},
  publisher={Oxford University Press}
}

@article{baym2015connect,
  title={Connect with your audience! The relational labor of connection},
  author={Baym, Nancy K},
  journal={The communication review},
  volume={18},
  number={1},
  pages={14--22},
  year={2015},
  publisher={Taylor \& Francis}
}

@misc{ohchrnduapa,
  title={The Impact of Counter-Terrorism Measures on Civil Society and Civic Space:The Indian (ab)use of counter-terrorism law Unlawful Activities Prevention Act in Kashmir},
  author={Office of the United Nations High Commissioner for Human Rights},
  howpublished={\url{https://www.ohchr.org/sites/default/files/documents/issues/terrorism/sr/cfis/cfi-gs-impact-ct-measures/subm-global-study-impact-cso-justice-all.pdf}},
  year={n.d.},
  note={[Accessed: March 28, 2026]}
}

@article{abidin2016visibility,
  title={Visibility labour: Engaging with Influencers’ fashion brands and\# OOTD advertorial campaigns on Instagram},
  author={Abidin, Crystal},
  journal={Media International Australia},
  volume={161},
  number={1},
  pages={86--100},
  year={2016},
  publisher={Sage Publications Sage UK: London, England}
}

@inproceedings{eslami2016first,
  title={First I" like" it, then I hide it: Folk Theories of Social Feeds},
  author={Eslami, Motahhare and Karahalios, Karrie and Sandvig, Christian and Vaccaro, Kristen and Rickman, Aimee and Hamilton, Kevin and Kirlik, Alex},
  booktitle={Proceedings of the 2016 cHI conference on human factors in computing systems},
  pages={2371--2382},
  year={2016}
}

@inproceedings{devito2017algorithms,
  title={" Algorithms ruin everything" \# RIPTwitter, Folk Theories, and Resistance to Algorithmic Change in Social Media},
  author={DeVito, Michael Ann and Gergle, Darren and Birnholtz, Jeremy},
  booktitle={Proceedings of the 2017 CHI conference on human factors in computing systems},
  pages={3163--3174},
  year={2017}
}

@article{marwick2017media,
  title={Media manipulation and disinformation online},
  author={Marwick, Alice and Lewis, Rebecca},
  journal={New York: Data \& Society Research Institute},
  volume={359},
  pages={1146--1151},
  year={2017}
}

@article{schauer1978fear,
  title={Fear, risk and the first amendment: Unraveling the chilling effect},
  author={Schauer, Frederick},
  journal={BUL rev.},
  volume={58},
  pages={685},
  year={1978},
  publisher={HeinOnline}
}

@article{metzger2010social,
  title={Social and heuristic approaches to credibility evaluation online},
  author={Metzger, Miriam J and Flanagin, Andrew J and Medders, Ryan B},
  journal={Journal of communication},
  volume={60},
  number={3},
  pages={413--439},
  year={2010},
  publisher={Oxford University Press}
}

@inproceedings{niu2023building,
  title={Building credibility, trust, and safety on video-sharing platforms},
  author={Niu, Shuo and Lu, Zhicong and Zhang, Amy X and Cai, Jie and Griggio, Carla F and Heuer, Hendrik},
  booktitle={Extended Abstracts of the 2023 CHI Conference on Human Factors in Computing Systems},
  pages={1--7},
  year={2023}
}

@inproceedings{haimson2016digital,
  title={Digital footprints and changing networks during online identity transitions},
  author={Haimson, Oliver L and Brubaker, Jed R and Dombrowski, Lynn and Hayes, Gillian R},
  booktitle={Proceedings of the 2016 CHI Conference on human factors in computing systems},
  pages={2895--2907},
  year={2016}
}

@book{marshall2006bengal,
  title={Bengal: The British Bridgehead: Eastern India 1740-1828},
  author={Marshall, Peter James},
  volume={2},
  year={2006},
  publisher={Cambridge University Press}
}

@misc{hrw2025india,
  title={India: Hundreds of Muslims Unlawfully Expelled to Bangladesh},
  author={Human Rights Watch},
  howpublished={\url{https://www.hrw.org/news/2025/07/23/india-hundreds-of-muslims-unlawfully-expelled-to-bangladesh}},
  year={2025},
  note={[Accessed: March 28, 2026]}
}

@misc{maitra2026india,
  title={India Is Using AI to Police Identity and Expel Minorities},
  author={Maitra, Suvradip},
  howpublished={\url{https://www.techpolicy.press/india-is-using-ai-to-police-identity-and-expel-minorities/}},
  year={2026},
  note={[Accessed: March 28, 2026]}
}

@misc{youtubendmulti,
  title={Multi-Channel Network (MCN) overview for YouTube Creators},
  author={YouTube Help},
  howpublished={\url{https://support.google.com/youtube/answer/2737059}},
  year={n.d.},
  note={[Accessed: March 31, 2026]}
}

@article{simpson2022hey,
  title={" Hey, Can You Add Captions?": The Critical Infrastructuring Practices of Neurodiverse People on TikTok},
  author={Simpson, Ellen and Dalal, Samantha and Semaan, Bryan},
  journal={arXiv preprint arXiv:2212.06204},
  year={2022}
}

@article{star1999ethnography,
  title={The ethnography of infrastructure},
  author={Star, Susan Leigh},
  journal={American behavioral scientist},
  volume={43},
  number={3},
  pages={377--391},
  year={1999},
  publisher={Sage Publications, Inc.}
}

@article{plantin2018infrastructure,
  title={Infrastructure studies meet platform studies in the age of Google and Facebook},
  author={Plantin, Jean-Christophe and Lagoze, Carl and Edwards, Paul N and Sandvig, Christian},
  journal={New media \& society},
  volume={20},
  number={1},
  pages={293--310},
  year={2018},
  publisher={Sage Publications Sage UK: London, England}
}

@misc{ahmed2025pluralism,
  title={Pluralism instead of secularism: theft in the house of ideas},
  author={Ahmed, Mohiuddin},
  howpublished={\url{https://www.prothomalo.com/opinion/column/jqlmyngmwu}},
  year={2025},
  note={[Accessed: July 4, 2026]}
}

@article{chen2014economic,
  title={An Economic Analysis of Online Advertising Using Behavioral Targeting1},
  author={Chen, Jianqing and Stallaert, Jan},
  journal={Mis Quarterly},
  volume={38},
  number={2},
  pages={429--450},
  year={2014},
  publisher={Management Information Systems Research Center, University of Minnesota}
}

@inproceedings{zhang2015analyzing,
  title={Analyzing and modeling special offer campaigns in location-based social networks},
  author={Zhang, Ke and Pelechrinis, Konstantinos and Lappas, Theodoros},
  booktitle={Proceedings of the International AAAI Conference on Web and Social Media},
  volume={9},
  number={1},
  pages={543--552},
  year={2015}
}

@article{aghapouri2020towards,
  title={Towards pluralistic and grassroots national identity: a study of national identity representation by the Kurdish diaspora on social media},
  author={Aghapouri, Jiyar Hossein},
  journal={National Identities},
  volume={22},
  number={2},
  pages={173--192},
  year={2020},
  publisher={Taylor \& Francis}
}

@article{fischer202114,
  title={14. Reconfiguring the Kurdish Nation on YouTube: Spatial Imaginations, Revolutionary Lyrics, and Colonial Knowledge},
  author={Fischer-Tahir, Andrea},
  journal={Media and Mapping Practices in the Middle East and North Africa},
  pages={299},
  year={2021}
}

\end{document}